\documentclass[%
 reprint,
superscriptaddress,
 amsmath,amssymb,
prl
]{revtex4-2}

\usepackage{graphicx}
\usepackage{dcolumn}
\usepackage{bm}
\usepackage[colorlinks,linkcolor=blue, urlcolor=blue, anchorcolor=blue, citecolor=blue]{hyperref}
\usepackage{subfigure}
\usepackage{float}

\begin{document}

\preprint{APS/123-QED}

\title{Instantaneous indirect measurement principle in quantum mechanics}

\author{Wangjun Lu}
\email{wjlu1227@zju.edu.cn}
\affiliation{Department of Maths and Physics, Hunan Institute of Engineering, Xiangtan 411104, China}
\affiliation{Zhejiang Institute of Modern Physics, Department of Physics, Zhejiang University, Hangzhou 310027, China}

\author{Xingyu Zhang}
\author{Lei Shao}
\author{Zhucheng Zhang}
\author{Jie Chen}
\author{Rui Zhang}
\author{Shaojie Xiong}
\author{Liyao Zhan}
\affiliation{Zhejiang Institute of Modern Physics, Department of Physics, Zhejiang University, Hangzhou 310027, China}

\author{Xiaoguang Wang}%
\email{xgwang1208@zju.edu.cn}
\affiliation{Zhejiang Institute of Modern Physics, Department of Physics, Zhejiang University, Hangzhou 310027, China}
\date{\today}

\begin{abstract}
In quantum systems, the measurement of operators and the measurement of the quantum states of the system are very challenging tasks. In this Letter, we propose a method to obtain the average value of one operator in a certain state by measuring the instantaneous change of the average value of another operator with the assistance of a known reference state. We refer to this measurement method as the instantaneous indirect measurement method. By studying the application of this method to some typical models, we find that this measurement can be applied to the measurement of an arbitrary state of a quantum system. Furthermore, for the system to be measured, we find that such measurement neither significantly affects the wave function of the system nor causes wave function collapse of the system. Also, our study shows that when two independent systems are coupled, the information mapping between them is done instantaneously. Finally, we discuss applying this measurement method to the measurement of quantum Fisher information, which quantizes the limited accuracy of estimating a parameter from a quantum state.
\end{abstract}

\maketitle


 
\textit{Introduction}.---In quantum mechanics, how to realize the measurement of an observable corresponding to a Hermitian operator in a certain state is an important research content \cite{braginsky1995quantum,ALLAHVERDYAN20131}. It is also a central problem in quantum physics to infer the quantum state of a system from the measurement results of different observables \cite{leonhardt1997measuring, RevModPhys.81.299}. A well-known method for implementing the measurement of the quadrature operators $\hat{X}(\theta)$ is the balanced homodyne detection method based on beam splitters \cite{RevModPhys.81.299}. Here, $\hat{X}(\theta)=\cos(\theta)\hat{x}+\sin(\theta)\hat{p}$ is an arbitrary linear superposition of the coordinate operator $\hat{x}$ and the momentum operator $\hat{p}$, where $\theta$ is a rotation angle. Theoretically, we can measure $\hat{X}(\theta)$ by changing the phase of the local coherent optical field in the balanced homodyne detection. Moreover, much theoretical and experimental work exists in the direction of measurement of operators \cite{PhysRevLett.85.1416, ASSmann:10, PhysRevA.76.013829, PhysRevA.82.043804, PhysRevLett.106.243601, Huang:16, PhysRevA.99.013839, settembrini2022detection,PhysRevLett.116.090801,PhysRevLett.127.260501,PhysRevResearch.3.043122}, such as measuring the correlation function of the optical field through quadrature measurements \cite{PhysRevA.82.043804}, single detectors with streak cameras \cite{ASSmann:10}, and random phase modulation \cite{Huang:16}. There is also quantum Fisher information extraction by random measurement methods \cite{PhysRevResearch.3.043122}.

Of course, we can also infer the state of the system from the measurement results of some operators, and this method of achieving quantum state reconstruction by some measurements is called quantum state tomography \cite{RevModPhys.81.299}.
Inspired by the theoretical work of Vogel and Risken \cite{PhysRevA.40.2847}, Smithey et al. used the balanced homodyne detection method to measure a set of probability densities for the quadrature amplitudes of a squeezed state and a vacuum state \cite{PhysRevLett.70.1244}. The Wigner functions of these two states were then reconstructed using the inverse Radon transform method. Another method of quantum superposition state reconstruction is quantum state holography, which is based on mixing the unknown state to be measured with a known reference state \cite{PhysRevLett.80.1418}. Various experimental realizations of quantum state holography have been reported \cite{weinacht1999controlling, huismans2011time, PhysRevA.101.013430}.

\begin{figure*}[ht]
\setlength{\abovecaptionskip}{0.cm}
\setlength{\belowcaptionskip}{-0.cm}
\centering
\includegraphics[width=18cm,height=9cm]{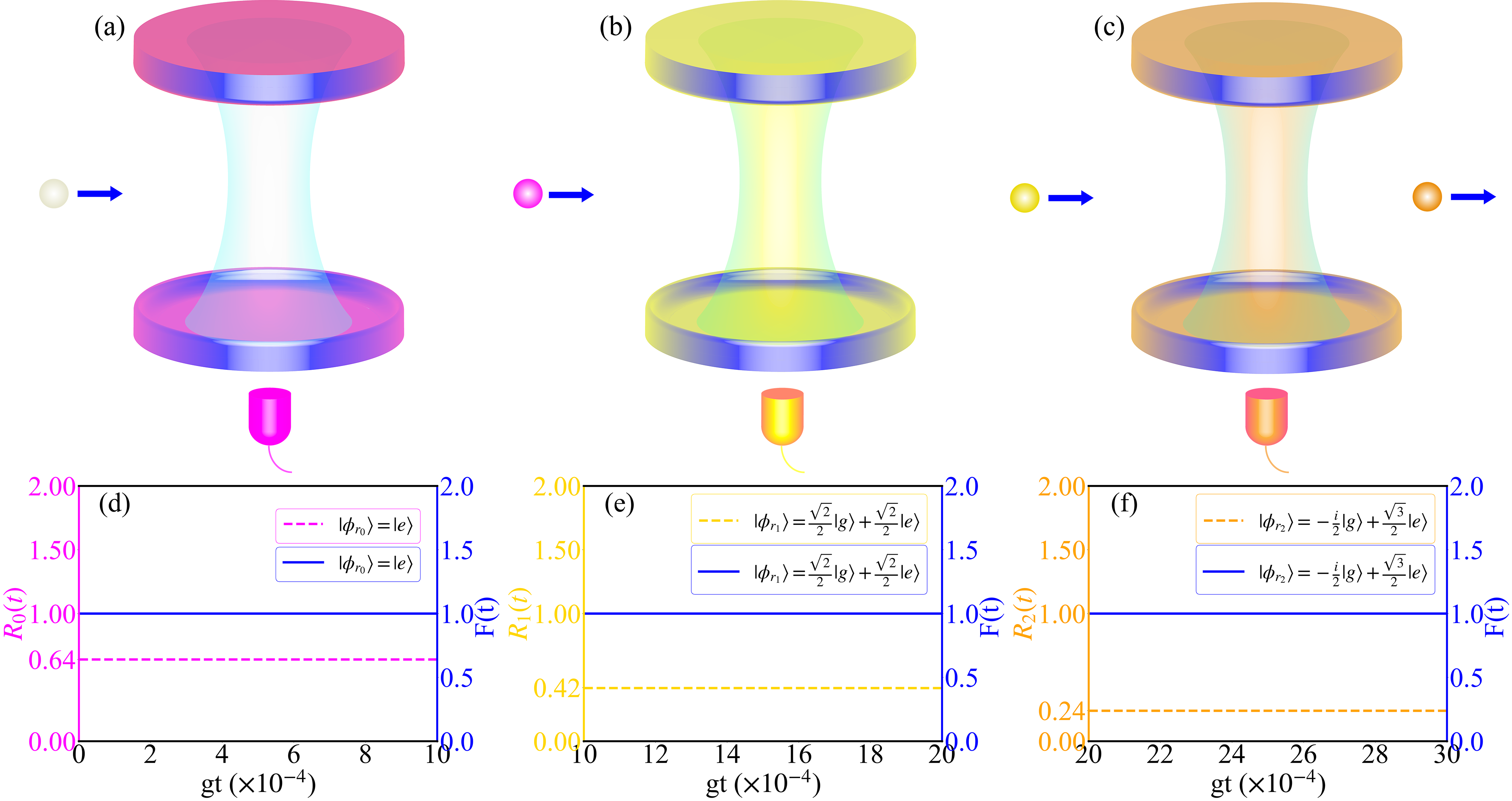}
\caption{(a), (b), (c) are schematic diagrams showing the process of obtaining the initial state information of a two-level atom by continuously measuring the average number of photons in three different cavity fields. The initial states of the cavity field in (a), (b), and (c) are the vacuum state $|N=0\rangle$, Coherent state $|\alpha=i\rangle$, and coherent state $|\alpha=1\rangle$. The state of the atom to be measured is $|\phi_{a}(0)\rangle=0.6|g\rangle+0.8e^{i\pi/6}|e\rangle$. (d), (e), and (f), respectively, show the changes of $R_{0}(t)$, $R_{1}(t)$, and $R_{2}(t)$ over time $t$, where $R_{0}(t)=\frac{\Delta\langle\hat{a}^{\dagger}\hat{a}\rangle(t,|\psi_{a}(0)\rangle)}{\Delta\langle\hat{a}^{\dagger}\hat{a}\rangle(t,|\psi_{r_{0}}\rangle)}\langle\psi_{r_{0}}|\hat{\sigma}_{+}\hat{\sigma}_{-}|\psi_{r_{0}}\rangle$, $R_{1}(t)=-0.5\frac{\Delta\langle\hat{a}^{\dagger}\hat{a}\rangle(t,|\psi_{a}(t_{1})\rangle)}{\Delta\langle\hat{a}^{\dagger}\hat{a}\rangle(t,|\psi_{r_{1}}\rangle)}\langle\psi_{r_{1}}|i(\hat{\sigma}_{+}\hat{a}-\hat{a}^{\dagger}\hat{\sigma}_{-})|\psi_{r_{1}}\rangle$, and $R_{2}(t)=0.5\frac{\Delta\langle\hat{a}^{\dagger}\hat{a}\rangle(t,|\psi_{a}(t_{2})\rangle)}{\Delta\langle\hat{a}^{\dagger}\hat{a}\rangle(t,|\psi_{r_{2}}\rangle)}\langle\psi_{r_{2}}|i(\hat{\sigma}_{+}\hat{a}-\hat{a}^{\dagger}\hat{\sigma}_{-})|\psi_{r_{2}}\rangle$. And $|\psi_{a}(0)\rangle =|N=0\rangle\otimes|\phi_{a}(0)\rangle  $, $|\psi_{a}(t_{1})\rangle=e^{-i\hat{H}_{JC}t_{1}}|\alpha=i\rangle\otimes|\phi_{a}(0)\rangle$, $|\psi_{a}(t_{2})\rangle=e^{-i\hat{H}_{JC}t_{2}}|\alpha=1\rangle\otimes|\phi_{a}(0)\rangle$, $|\psi_{r_{0}}\rangle=|N=0\rangle\otimes|\phi_{r_{0}}\rangle$, $|\psi_{r_{1}}\rangle=|\alpha=i\rangle\otimes|\phi_{r_{1}}\rangle$, $|\psi_{r_{2}}\rangle=|\alpha=1\rangle\otimes|\phi_{r_{2}}\rangle$, where $t_{1}=0.001g^{-1}$ and $t_{2}=0.002g^{-1}$. At the same time, we plot the fidelity $F(t)$ of the atomic state relative to the initial atomic state as a function of time $t$ in (d), (e) , and (f). Here, the three known reference atomic states used when measuring the excited state population probabilities of the atom and the real and imaginary parts of the off-diagonal elements of the atomic state density matrix are shown in the legends of (d), (e), and (f). The values of other parameters are $w_{a}=1$, $w_{0}=w_{a}$, $U=0.1w_{a}$, and $\gamma=0.2w_{a}$.}
\label{fig1}
\end{figure*}

Recently, wave function measurements of quantum pure states based on weak measurement methods have attracted a lot of research interest \cite{lundeen2011direct,PhysRevA.84.052107,PhysRevA.86.052110,PhysRevLett.108.070402,PhysRevLett.117.120401,PhysRevLett.117.170402,PhysRevLett.113.090402,PhysRevLett.123.150402,PhysRevLett.116.040502}. This measurement method is called the direct measurement method, which was proposed by Lundeen et al \cite{lundeen2011direct}. Here,  ``direct" means that no complicated measurements and calculations are needed after the measurement. In weak measurements, the weak value can be a complex number whose imaginary and real parts can be obtained by measuring a pair of conjugate quantities. The direct measurement method relies on the sequential measurement of two complementary variables of the system. To reduce the perturbation of the wave function caused by the first measurement, a weak measurement of the first variable is required, followed by a normal measurement of the second complementary variable \cite{lundeen2011direct}. The state reconstruction method for quantum systems by weak measurements is then extended successively to continuous-variable quantum systems \cite{PhysRevA.86.052110} and mixed state quantum systems \cite{PhysRevLett.108.070402}. Mirhossein et al. combined the advantages of direct measurement methods with a compressive sensing technique and found that the wave functions of high-dimensional states can be estimated with high fidelity using a much smaller number of measurements than standard direct measurement methods \cite{PhysRevLett.113.090402}. Pan et al. demonstrated a direct measurement of nonlocal wave functions using the modular values method \cite{PhysRevLett.123.150402}. However, it is worth noting that, unlike phase-space measurements, reconstructing a completely unknown quantum state is not always possible using weak measurements \cite{PhysRevA.84.052107}. Vallone and Dequal found that strong measurements,  stronger coupling between the system and the measurement device, allow better direct measurements of the quantum wave function and also demonstrated that weak measurements are not necessary for direct measurements of the wave function \cite{PhysRevLett.116.040502}.

While studying the Tavis-Cummings quantum battery, we discovered an interesting phenomenon with the formula 
$\lim_{t\rightarrow0}\frac{F(M_{1},t)}{F(M_{2},t)}=\frac{M_{1}}{M_{2}}$  \cite{PhysRevA.104.043706}. Here $F(M_{1},t)$ and $F(M_{2},t)$ are the energy changes of the atoms when the initial average photon number of the cavity field are $M_{1}$ and $M_{2}$, respectively. After analyzing the reason behind it, we proved the following instantaneous indirect measurement principle (IIMP) in quantum mechanics.

\textit{The IIMP in quantum mechanics}.---We consider a Hamiltonian $\hat{H}$ and its initial state $\left|\Psi(0)\right\rangle$ , then the state at time $t$ is ($\hbar=1$)
\begin{equation}
\left|\Psi(t)\right\rangle=\exp(-i\hat{H}t)\left|\Psi(0)\right\rangle.   \label{1}
\end{equation}
The difference of the mean values of a Hermitian operator $\hat{A}$ of the system between the initial state and the state at time $t$ is
\begin{equation}
\Delta\langle \hat{A}\rangle(t, \left|\Psi(0)\right\rangle ) =\left\langle \Psi(t)\right|\hat{A}\left|\Psi(t)\right\rangle -\left\langle \Psi(0)\right|\hat{A}\left|\Psi(0)\right\rangle. \label{2}
\end{equation}
When the system evolves with another known reference initial state $\left|\Psi_{r}(0)\right\rangle$, then
\begin{equation}  
\Delta\langle \hat{A}\rangle(t,\left|\Psi_{r}(0)\right\rangle ) =\left\langle \Psi_{r}(t)\right|\hat{A}\left|\Psi_{r}(t)\right\rangle -\left\langle \Psi_{r}(0)\right|\hat{A}\left|\Psi_{r}(0)\right\rangle.      \label{3}
\end{equation}

When $t\rightarrow0$, the ratio of the change in the mean value of the operator $\hat{A}$ for different initial states is \cite{Suppl}
\begin{equation}
\lim_{t\rightarrow0}\frac{\Delta\langle \hat{A}\rangle(t,\left|\Psi(0)\right\rangle ) }{\Delta\langle \hat{A}\rangle(t,\left|\Psi_{r}(0)\right\rangle ) }=\frac{\left\langle \Psi(0)\right|(i\hat{H})^{\times n}(\hat{A})\left|\Psi(0)\right\rangle }{\left\langle \Psi_{r}(0)\right|(i\hat{H})^{\times n}(\hat{A})\left|\Psi_{r}(0)\right\rangle }, \label{4}
\end{equation}
where the superoperator $(i\hat{H})^{\times n}(\cdot)$ denotes a $n$th order nested commutator operation, and
\begin{equation}
(i\hat{H})^{\times n}(\hat{A})=[\underset{n}{\underbrace{i\hat{H},\cdots[i\hat{H},[i\hat{H}}},\hat{A}]]],  \label{5}
\end{equation}
where $n$ is determined by when the $n$th order derivative of $\Delta\langle \hat{A}\rangle(t, \left|\Psi(0)\right\rangle ) $  with respect to time $t$ is not zero for the first time. It is worth noting that in Eq.~(\ref{4}), the state $\left|\Psi_{r}(0)\right\rangle$ is known, so we can obtain the average value of the operator $(i\hat{H})^{\times n}(\hat{A})$ in the initial state by measuring the change of the operator $\hat{A}$ in a short time, and eventually, we can obtain the information of the state $\left|\Psi(0)\right\rangle$, i.e.,
\begin{eqnarray}
&&\left\langle \Psi(0)\right|(i\hat{H})^{\times n}(\hat{A})\left|\Psi(0)\right\rangle\nonumber\\
&=&\left\langle \Psi_{r}(0)\right|(i\hat{H})^{\times n}(\hat{A})\left|\Psi_{r}(0)\right\rangle \lim_{t\rightarrow0}\frac{\Delta\langle \hat{A}\rangle(t,\left|\Psi(0)\right\rangle ) }{\Delta\langle \hat{A}\rangle(t,\left|\Psi_{r}(0)\right\rangle ) }.   \label{6}
\end{eqnarray} 
Similar to the above, when the system evolves with an unknown mixed state $\hat{\rho}(0)$ and a known reference mixed state $\hat{\rho}_{r}(0)$, respectively, we can obtain the following equation
\begin{eqnarray}
&&{\rm Tr}[(i\hat{H})^{\times n}(\hat{A})\hat{\rho}(0)]\nonumber\\
&=&{\rm Tr}[(i\hat{H})^{\times n}(\hat{A})\hat{\rho}_{r}(0)]\lim_{t\rightarrow0}\frac{\Delta\langle \hat{A}\rangle (t,\hat{\rho}(0))}{\Delta\langle \hat{A}\rangle (t,\hat{\rho}_{r}(0))}.   \label{7}
\end{eqnarray}
In particular, if $\langle\hat{A}\rangle$ and $\langle(i\hat{H})^{\times n}(\hat{A})\rangle$ belong to different subsystems in the composite system, we can obtain the information of the state of the one subsystem by measuring the change of a quantity in another subsystem. Such a measurement can be maximized without affecting the target system. Of course, we can also pick other observable measurements of the system to measure, and we will not continue to expand the discussion here. It is noteworthy that this instantaneous indirect measurement method we propose is able to measure any state of a system at any time during its dynamical evolution, and this measurement has almost no effect on the system to be measured. Unlike the known reference states utilized in Ref. \cite{PhysRevLett.80.1418}, which need to be involved in the dynamics of the system to be measured, the reference states we utilize do not need to be involved in the dynamics of the system to be measured. Therefore, the influence of the reference state on the system to be measured is thus avoided.

\textit{Applications of the IIMP to some specific models}.--- We considering a extended $p$-photon Jaynes-Cummings model, its Hamiltonian is
\begin{equation}
\hat{H}_{JC}=\omega_{a}\text{\ensuremath{\hat{a}^{\dagger}}}\hat{a}+\frac{\omega_{0}}{2}\hat{\sigma}_{z}+g(\hat{a}^{\dagger p}\hat{\sigma}_{-}+\hat{\sigma}_{+}\hat{a}^{p})+\frac{U}{2}\hat{a}^{\dagger2}\hat{a}^{2}+\gamma\text{\ensuremath{\hat{a}^{\dagger}}}\hat{a}\hat{\sigma}_{z} ,   \label{8}
\end{equation}
where $p$ is an integer greater than or equal to $1$. $\omega_{a}$ and $\omega_{0}$ are the eigenfrequency of the optical cavity and the transition frequency of the two-level atom, respectively. $g$ is the dipole interaction strength between the cavity field and the single atom, $U$ is the Kerr-type nonlinear interaction strength of the optical field, and $\gamma$ is the dispersion interaction between the cavity field and the single atom. $\hat{a}$ and $\hat{a}^{\dagger}$ are, respectively, the annihilation and creation operators of the single-mode cavity field, they satisfy the usual bosonic commutation relation $[\hat{a},\hat{a}^{\dagger}]=1$. $\hat{\sigma}_{x,y,z}$ are the usual Pauli operators, and $\hat{\sigma}_{\pm}=\frac{1}{2}(\hat{\sigma}_{x}\pm i\hat{\sigma}_{y})$.

First, we choose the average photon number of the cavity field as the object of observation and measure the unknown initial state of a two-level atom by the variation of the average photon number of a known optical field over a short period of time with the assistance of a known reference state. In the following, we first analyze how the diagonal and non-diagonal elements of the density matrix of a two-level atom can be measured separately by different known cavity fields with the aid of a reference state. Then, a concrete example is used to demonstrate how the information about the unkonwn state of the atom can be completely obtained through the change of the average photon number of the three known cavity fields in a short time under the actual dynamics. At the same time, we find that such instantaneous indirect measurements have almost no effect on the state of the atom.
We choose the photon number of the cavity field as the measurement object. According to our principle of instantaneous indirect measurement, when the initial state of the cavity field is a vacuum state $\left|\varphi\right\rangle=\left|0\right\rangle$ and the initial state of the atom to be measured is an unknown state $\left|\phi_{a}\right\rangle$, we can obtain the following expression  \cite{Suppl}
\begin{eqnarray}
&&\left\langle \phi_{a}\right|\hat{\sigma}_{+}\hat{\sigma}_{-}\left|\phi_{a}\right\rangle \nonumber \\ &=&\lim_{t\rightarrow0}\frac{\Delta\left\langle \text{\ensuremath{\hat{a}^{\dagger}}}\hat{a}\right\rangle\left(t,\left|\psi(0)\right\rangle \right) }{\Delta\left\langle \text{\ensuremath{\hat{a}^{\dagger}}}\hat{a}\right\rangle\text{\ensuremath{\left(t,\left|\psi_{r_{0}}(0)\right\rangle \right)}} }\left\langle \phi_{r_{0}}\right|\hat{\sigma}_{+}\hat{\sigma}_{-}\left|\phi_{r_{0}}\right\rangle,  \label{9}
\end{eqnarray}
where $\left|\psi(0)\right\rangle=\left|0\right\rangle\otimes\left|\phi_{a}\right\rangle$ and $\left|\psi_{r_{0}}(0)\right\rangle=\left|0\right\rangle\otimes\left|\phi_{r_{0}}\right\rangle$. $\left|\phi_{r_{0}}\right\rangle$ is a known reference state of the atomic system, where its choice is arbitrary except that it makes the average value of $\hat{\sigma}_{+}\hat{\sigma}_{-}$ equal to zero. In Fig.~\ref{fig1}(a), we present a schematic diagram of measuring the probability population of a two-level atom by measuring the change in the average photon number of the cavity field. For an atomic state $\left|\phi_{a}\right\rangle=0.6\left|g\right\rangle+0.8e^{i\frac{\pi}{6}}\left|e\right\rangle$ to be measured, when we choose the initial state of the cavity field to be the vacuum state $\left|0\right\rangle$ and the reference state of the atom to be the excited state $\left|\phi_{r_{0}}\right\rangle=\left|e\right\rangle$, we plot the mean photon number of the cavity field and the fidelity of the atomic state as a function of time with the aid of a known initial state, respectively, in Fig.~\ref{fig1}(d). This fidelity is defined as follows
\begin{equation}
F(t)=\left|\left\langle\phi_{a}(0)\right|e^{-i\hat{H}_{JC}t}\left|\phi_{a}(0)\right\rangle\right|^{2}. \label{10}
\end{equation}
We find that with the aid of this initial reference state, the probability population of the atom in the excited state can be obtained by measuring the instantaneous change in the average photon number of the cavity field. And we found that the fidelity between the atomic state and the initial state at any time is always close to 1; that is to say, this instantaneous indirect measurement process has almost no effect on the initial state of the atom.

Then, when we choose the initial state of the cavity field to be the coherent state $\left|\alpha\right\rangle$, we can obtain  \cite{Suppl}
\begin{eqnarray}
&&\left\langle \phi_{a}\right|\left(\hat{\sigma}_{+}\alpha-\alpha^{*}\hat{\sigma}_{-}\right)\left|\phi_{a}\right\rangle \hfill\nonumber\\
&=&\lim_{t\rightarrow0}\!\frac{\Delta\left\langle \text{\ensuremath{\hat{a}^{\dagger}}}\hat{a}\right\rangle\left(t,\left|\psi(0)\right\rangle \right) }{\Delta\left\langle \text{\ensuremath{\hat{a}^{\dagger}}}\hat{a}\right\rangle\text{\ensuremath{\left(t,\left|\psi_{r}(0)\right\rangle \right)}} }\!\left\langle \phi_{r}\right|\!\left(\hat{\sigma}_{+}\alpha-\alpha^{*}\hat{\sigma}_{-}\right)\!|\phi_{r}\rangle \hfill ,\label{11}
\end{eqnarray}
where $\left|\psi(0)\right\rangle=\left|\alpha\right\rangle\otimes\left|\phi_{a}\right\rangle$, $\left|\psi_{r}(0)\right\rangle=\left|\alpha\right\rangle\otimes\left|\phi_{r}\right\rangle$, and $\left|\phi_{r}\right\rangle$ are shown in the legends of Fig.~\ref{fig1}(e) and  Fig.~\ref{fig1}(f) when $\alpha$ takes different values. Obviously, when we choose $\alpha=i$ or $\alpha=1$, we can obtain the real or imaginary part of the non-diagonal elements of the two-level atomic density matrix by the instantaneous change of the average photon number of the cavity field. Therefore, in order to measure the real and imaginary parts of the off-diagonal elements of the atomic state density matrix after the atom pass through the first vacuum field, we prepare the cavity fields in Fig.~\ref{fig1}(b) and Fig.~\ref{fig1}(c) in the coherent states $\left|\alpha=i\right\rangle$ and $\left|\alpha=1\right\rangle$, respectively.  Similar to above, we plot both the average photon number of the cavity field and the fidelity of the atomic state with the aid of different known reference states as a function of time in both Fig.~\ref{fig1}(e) and Fig.~\ref{fig1}(f). We found that the real and imaginary parts of the off-diagonal elements of the atomic state density matrix can be obtained by measuring the change of the average photon number of the cavity field with the aid of the reference state. Eventually, we obtained the initial density matrix of the atom by measuring the variation of the average photon number in three different cavity fields, i.e., $\hat{\rho}_{atom}=0.64\left|e\right\rangle\left\langle e\right|+0.36\left|g\right\rangle\left\langle g\right|+(0.42+0.24i)\left|e\right\rangle\left\langle g\right|+(0.42-0.24i)\left|g\right\rangle\left\langle e\right|$. Moreover, the fidelity of the atomic states shows that this instantaneous indirect measurement has almost no effect on the states of the atom.

\begin{figure}[t!]
\setlength{\abovecaptionskip}{0.cm}
\setlength{\belowcaptionskip}{-0.cm}
\centering
\includegraphics[width=8.5cm,height=10cm]{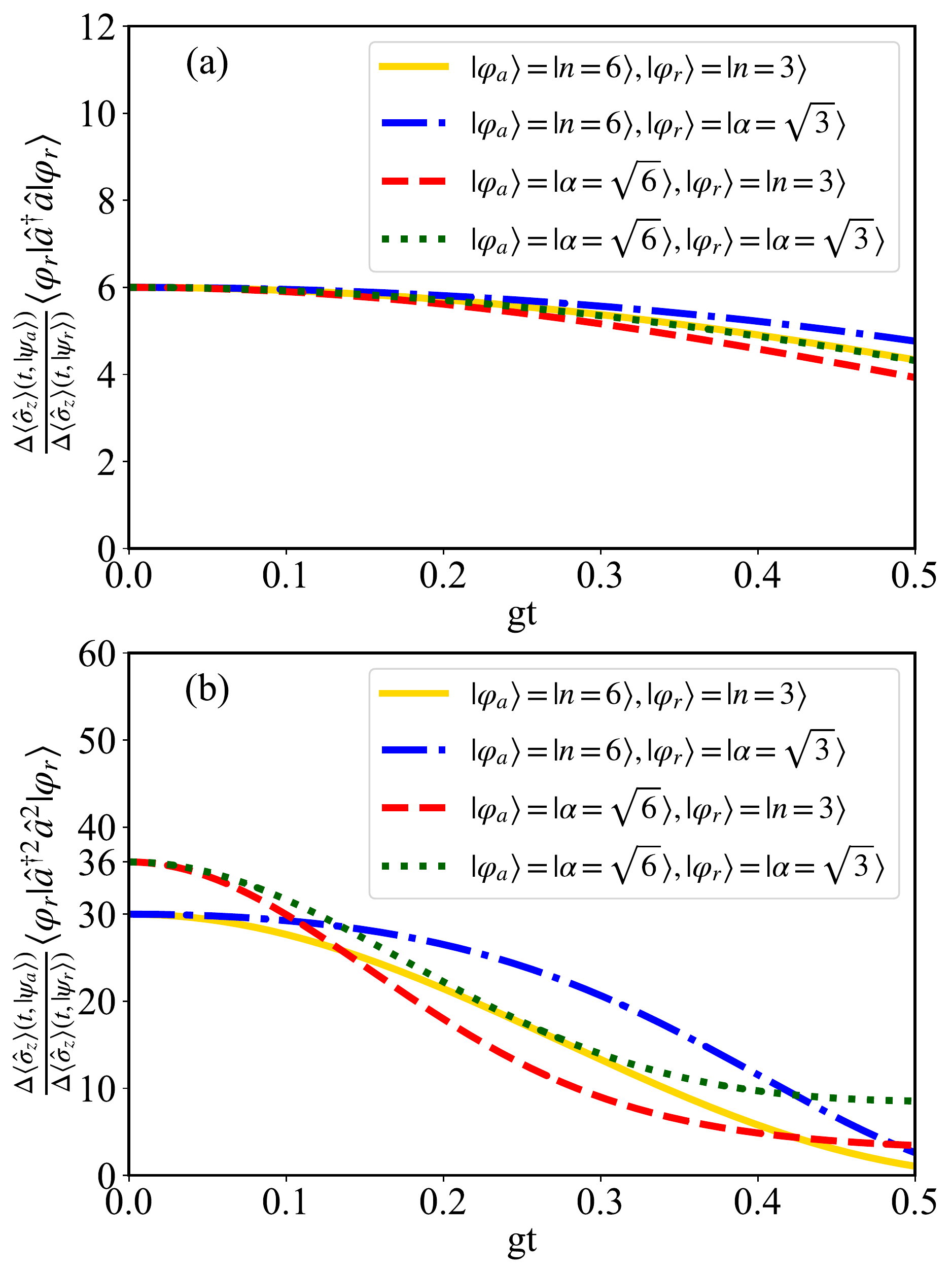}
\caption{ (a) and (b) show the relative change of atomic energy versus time in the single-photon J-C model and the two-photon J-C model, respectively. Here, $\left|n\right\rangle$ and $\left|\alpha\right\rangle$ denote respectively the number and coherent states, and the other parameters are the same as in Fig. \ref{fig1}.}
\label{fig2}
\end{figure}

Second, we choose $\hat{\sigma}_{z}$ as the measurement object to obtain the information of the cavity field. When the cavity field is in an arbitrary state $|\varphi_{a}\rangle$, if the atom is initially in the ground state, i.e., $\left|\psi(0)\right\rangle =|\varphi_{a}\rangle\otimes|g\rangle$, then we can obtain \cite{Suppl}
\begin{equation}
\left\langle \varphi_{a}\right|\hat{a}^{\dagger p}\hat{a}^{p}|\varphi_{a}\rangle=\lim_{t\rightarrow0}\frac{\Delta\left\langle \hat{\sigma}_{z}\right\rangle(t,\left|\psi(0)\right\rangle ) }{\Delta\left\langle \hat{\sigma}_{z}\right\rangle(t,\left|\psi_{r}(0)\right\rangle ) }\left\langle \varphi_{r}\right|\hat{a}^{\dagger p}\hat{a}^{p}|\varphi_{r}\rangle,  \label{12}
\end{equation} 
where $|\psi_{r}\rangle=|\varphi_{r}\rangle\otimes|g\rangle$ is a known reference state. Obviously, we can get the $p$-order correlation function of the unknown state $\varphi_{a}$ of the cavity field by measuring the instantaneous change of the atomic energy. For example, when $p$ is equal to $1$ and $2$, respectively, we can obtain respectively the initial average photon number and the second-order correlation function of the unknown cavity field by measuring the energy change of the atom. In Fig. \ref{fig2}(a) and Fig. \ref{fig2}(b), we plot the energy change of the atom with the aid of the reference state as a function of time when $p=1$ and $p=2$, respectively. We find that the $p$-order correlation function of the cavity field can be obtained by measuring the energy change of the atom with the aid of the reference state. And it is found that the reference state selected as the coherent state is better than the number state under the same number of photons. Furthermore, when we prepare the initial state of the probe atom at $|\phi\rangle=(|g\rangle-ie^{i\theta}|e\rangle)/\sqrt{2}$, we can measure the quadrature operators  $\hat{X}(\theta)$ by measuring the energy change of the probe atom \cite{Suppl}. This indirect measurement method has almost no effect on the optical field to be measured compared to balanced homodyne detection.
  
We have studied how to measure the cavity field to obtain the information of the atom and how to measure the atom to get the information of the cavity field above, and we show that this instantaneous indirect measurement has almost no effect on the target system. At the same time, if the operator corresponding to the direct measurement quantity we choose commute with some quantities in Hamiltonian, these quantities will not affect the measurement result. For example, in the extended $p$-photon JC model, when the direct measurement quantity we choose is the average photon number of the cavity field or the energy of the atom, since $[\hat{a}^{\dagger}\hat{a}, \hat{a}^{\dagger p}\hat{a}^{p}]=[\hat{a}^{\dagger}\hat{a}, \hat{a}^{\dagger}\hat{a}\hat{\sigma}_{z}]=0$ or $[\hat{\sigma}_{z}, \hat{a}^{\dagger p}\hat{a}^{p}]=[\hat{\sigma}_{z}, \hat{a}^{\dagger}\hat{a}\hat{\sigma}_{z}]=0$, the frequency detuning between the cavity field and the atom, the nonlinear interaction between the photons and the photon-atom interaction do not affect the measurement results.

In the following, we study the instantaneous indirect measurement method used to measure the quantum Fisher information of the system. When a Hamiltonian $\hat{H}(\lambda)$ with parameter $\lambda$ evolves with an arbitrary initial state $|\Psi(0)\rangle$ and a known reference initial state $|\Psi_{r}(0)\rangle$, respectively, then we can obtain the following relation \cite{Suppl}
\begin{equation}
\lim_{t\rightarrow0}\frac{F_{\lambda}(t,\left|\Psi(0)\right\rangle )}{F_{\lambda}(t,\left|\Psi_{r}(0)\right\rangle )}=\frac{\left\langle \Psi(0)\right|\Delta\left[\frac{\partial\hat{H}(\lambda)}{\partial\lambda}\right]^{2}\left|\Psi(0)\right\rangle }{\left\langle \Psi_{r}(0)\right|\Delta\left[\frac{\partial\hat{H}(\lambda)}{\partial\lambda}\right]^{2}\left|\Psi_{r}(0)\right\rangle },   \label{13}
\end{equation}
where $\left\langle \Psi(0)\right|\Delta[\partial\hat{H}(\lambda)/\partial\lambda]^{2}\left|\Psi(0)\right\rangle$ and $\left\langle \Psi_{r}(0)\right|\Delta[\partial\hat{H}(\lambda)/\partial\lambda]^{2}\left|\Psi_{r}(0)\right\rangle $  are the variances of the Hermitian operator $\partial\hat{H}(\lambda)/\partial\lambda$ in the initial states $\left|\Psi(0)\right\rangle$ and a know reference state $\left|\Psi_{r}(0)\right\rangle$, respectively. 

For the extended JC model shown in Eq.~(\ref{8}), we choose the coupling strength $g$ as the parameter to be measured, and the initial state of the system is $|\psi(0)\rangle=|\varphi_{a}(0)\rangle\otimes|g\rangle$, where the cavity field is in an arbitrary state $|\varphi_{a}(0)\rangle$ and the atom is in the ground state $|g\rangle$. With the aid of a known reference state $|\psi_{r}(0)\rangle=|\varphi_{r}(0)\rangle\otimes|g\rangle$ ($|\varphi_{r}(0)\rangle$ is a known reference state of the cavity field), we can obtain
\begin{eqnarray}
\lim_{t\rightarrow0}\frac{F_{\lambda}(t,|\psi_{a}(0)\rangle )}{F_{\lambda}(t,|\psi_{r}(0)\rangle)}&=&\frac{\left\langle\varphi_{a}(0)\right|\hat{a}^{\dagger p}\hat{a}^{p}\left|\varphi_{a}(0)\right\rangle}{\left\langle\varphi_{r}(0)\right|\hat{a}^{\dagger p}\hat{a}^{p}\left|\varphi_{r}(0)\right\rangle} \nonumber\\ &=&\lim_{t\rightarrow0}\frac{\Delta\langle\hat{\sigma}_{z}\rangle(t,|\psi(0)\rangle)}{\Delta\langle\hat{\sigma}_{z}\rangle(t,|\varphi_{r}(0)\rangle)}. \label{14}
\end{eqnarray}
Thus, after an instantaneous evolution of the system, i.e., an evolution time $t_{0}$ close to 0, we can indirectly measure the quantum Fisher information for the parameter $g$ at the moment $t_{0}$ by measuring the instantaneous change in the energy of the atom with the aid of a known reference state $|\psi_{r}(0)\rangle$, i.e.,
\begin{equation}
F_{\lambda}(t_{0},|\psi_{a}(0)\rangle )\approx \frac{\Delta\langle\hat{\sigma}_{z}\rangle(t_{0},|\psi(0)\rangle)}{\Delta\langle\hat{\sigma}_{z}\rangle(t_{0},|\varphi_{r}(0)\rangle)}F_{\lambda}(t_{0},|\psi_{r}(0)\rangle). \label{15}
\end{equation}

\textit{Conclusion}.---In summary, we propose an instantaneous indirect measurement method. With the aid of a known reference state, we can obtain the initial value of one observable by measuring the instantaneous change of another observable, which shows that the mapping of information from one subsystem to another is done instantaneously. We have studied in some typical models how to measure the energy change of an atom to obtain information about the cavity field, and vice versa. We found that this indirect measurement minimizes the impact of the measurement on the system to be measured. In particular, this indirect measurement method is able to measure the state of the system to be measured at any moment with little effect on the dynamical evolution of the system. Finally, we discuss using this measurement method to measure the quantum Fisher information of the system.

W.J.Lu thanks Yan Liu, Xiongying Dai, Yonghe Deng, and Qiao Chen for valuable discussions. X.G.W was supported by the NSFC through Grant
No. 11935012 and No. 11875231, the National Key
Research and Development Program of China (Grants
No. 2017YFA0304202 and No. 2017YFA0205700).
W.J.L. was supported by the NSFC( Grants No.
11947069) and the Scientific Research Fund of Hunan Provincial Education Department( Grants No.
20C0495).
\nocite{*}

\bibliography{apssamp}

\end{document}


\preprint{APS/123-QED}

\title{Supplemental Material for ``Instantaneous indirect measurement principle in quantum mechanics" }

\author{Wangjun Lu}
\email{wjlu1227@zju.edu.cn}
\affiliation{Department of Maths and Physics, Hunan Institute of Engineering, Xiangtan 411104, China}
\affiliation{Zhejiang Institute of Modern Physics, Department of Physics, Zhejiang University, Hangzhou 310027, China}

\author{Xingyu Zhang}
\author{Lei Shao}
\author{Zhucheng Zhang}
\author{Jie Chen}
\author{Rui Zhang}
\author{Shaojie Xiong}
\author{Liyao Zhan}
\affiliation{Zhejiang Institute of Modern Physics, Department of Physics, Zhejiang University, Hangzhou 310027, China}

\author{Xiaoguang Wang}%
\email{xgwang1208@zju.edu.cn}
\affiliation{Zhejiang Institute of Modern Physics, Department of Physics, Zhejiang University, Hangzhou 310027, China}

\date{\today}

\maketitle

\tableofcontents

\section{A proof and applications of the instantaneous indirect measurement principle in quantum mechanics }
In this section, we first give a proof of the instantaneous indirect measurement principle (IIMP), then apply it to some specific models for discussion, and finally validate it against the actual dynamical evolution with each other.

\subsection{A proof of the instantaneous indirect measurement principle in quantum mechanics }
In this subsection, we first study the system whose initial state is pure, and later we study the system whose initial state is mixed. We consider a Hermitian Hamiltonian $\hat{H}$ and a initial state $\left|\Psi(0)\right\rangle$ , then the state at time $t$ is
\begin{equation}
\left|\Psi(t)\right\rangle =\exp(-i\hat{H}t)\left|\Psi(0)\right\rangle.   \tag{S1}\label{S1}
\end{equation}
The difference between the average value of a Hermitian operator $\hat{A}$ of the system over the initial and final states is
\begin{align}
	\Delta\langle \hat{A}\rangle(t,\left|\Psi(0)\right\rangle ) &=\left\langle \Psi(t)\right|\hat{A}\left|\Psi(t)\right\rangle -\left\langle \Psi(0)\right|\hat{A}\left|\Psi(0)\right\rangle \nonumber\\
	&=\left\langle \Psi(0)\right|\exp(i\hat{H}t)\hat{A}\exp(-i\hat{H}t)\left|\Psi(0)\right\rangle -\left\langle \Psi(0)\right|\hat{A}\left|\Psi(0)\right\rangle. \tag{S2}\label{S2}
\end{align}
When the system evolves in another initial state $\left|\Psi_{r}(0)\right\rangle$ (It is a known reference state), then
\begin{equation}
\Delta\langle \hat{A}\rangle(t,\left|\Psi_{r}(0)\right\rangle ) =\left\langle \Psi_{r}(0)\right|\exp(i\hat{H}t)\hat{A}\exp(-i\hat{H}t)\left|\Psi_{r}(0)\right\rangle -\left\langle \Psi_{r}(0)\right|\hat{A}\left|\Psi_{r}(0)\right\rangle. \tag{S3}\label{S3}
\end{equation}
When $t\rightarrow0$, the ratio of the change in the mean value of the operator $\hat{A}$ for different initial states is
\begin{equation}
\lim_{t\rightarrow0}\frac{\Delta\langle \hat{A}\rangle(t,\left|\Psi(0)\right\rangle ) }{\Delta\langle \hat{A}\rangle(t, \left|\Psi_{r}(0)\right\rangle ) }=\lim_{t\rightarrow0}\frac{\left\langle \Psi(0)\right|\exp(i\hat{H}t)\hat{A}\exp(-i\hat{H}t)\left|\Psi(0)\right\rangle -\left\langle \Psi(0)\right|\hat{A}\left|\Psi(0)\right\rangle}{\left\langle \Psi_{r}(0)\right|\exp(i\hat{H}t)\hat{A}\exp(-i\hat{H}t)\left|\Psi_{r}(0)\right\rangle -\left\langle \Psi_{r}(0)\right|\hat{A}\left|\Psi_{r}(0)\right\rangle}.\tag{S4}\label{S4}
\end{equation}
Since $\lim_{t\rightarrow0}\Delta\langle \hat{A}\rangle(t,\left|\Psi(0)\right\rangle ) =0$, $\lim_{t\rightarrow0}\Delta\langle \hat{A}\rangle(t, \left|\Psi_{r}(0)\right\rangle )=0$, and 
\begin{align}
	\lim_{t\rightarrow0}\frac{d^{n}\Delta\langle \hat{A}\rangle(t,\left|\Psi(0)\right\rangle ) }{dt^{n}}&=\lim_{t\rightarrow0}\left\langle \Psi(0)\right|\frac{d^{n}\left[\exp(i\hat{H}t)\hat{A}\exp(-i\hat{H}t)\right]}{dt^{n}}\left|\Psi(0)\right\rangle \nonumber\\
	&=\lim_{t\rightarrow0}\left\langle \Psi(0)\right|\frac{d^{n}}{dt^{n}}\Big[\hat{A}+t[i\hat{H},\hat{A}]+\frac{t^{2}}{2}[i\hat{H},[i\hat{H},\hat{A}]]+\frac{t^{3}}{3!}[i\hat{H},[i\hat{H},[i\hat{H},\hat{A}]]]+\cdots\nonumber\\
	&+\frac{t^{n}}{n!}[\underset{n}{\underbrace{i\hat{H},\cdots[i\hat{H},[i\hat{H}}},\hat{A}]]]+\cdots\Big]\left|\Psi(0)\right\rangle \nonumber\\
	&=\left\langle \Psi(0)\right|\Big[\underset{n}{\underbrace{i\hat{H},\cdots[i\hat{H},[i\hat{H}}},\hat{A}]]\Big]\left|\Psi(0)\right\rangle ,  \tag{S5}\label{S5} \\
	\lim_{t\rightarrow0}\frac{d^{n}\Delta\langle \hat{A}\rangle(t, \left|\Psi_{r}(0)\right\rangle ) }{dt^{n}}&=\left\langle \Psi_{r}(0)\right|\Big[\underset{n}{\underbrace{i\hat{H},\cdots[i\hat{H},[i\hat{H}}},\hat{A}]]\Big]\left|\Psi_{r}(0)\right\rangle ,  \tag{S6}\label{S6}
\end{align}
In the above, we used the Baker-Campbell-Hausdorff formula. Then, according to L'H\^{o}pital's rule, we can obtain
\begin{equation}
\lim_{t\rightarrow0}\frac{\Delta\langle \hat{A}\rangle(t,|\Psi(0)\rangle ) }{\Delta\langle \hat{A}\rangle(t,|\Psi_{r}(0)\rangle ) }=\frac{\left\langle \Psi(0)\right|[\underset{n}{\underbrace{i\hat{H},\cdots[i\hat{H},[i\hat{H}}},\hat{A}]]]\left|\Psi(0)\right\rangle }{\left\langle \Psi_{r}(0)\right|[\underset{n}{\underbrace{i\hat{H},\cdots[i\hat{H},[i\hat{H}}},\hat{A}]]]\left|\Psi_{r}(0)\right\rangle },  \tag{S7}\label{S7}
\end{equation}
where $n$ is determined by when the $n$th order derivative of $\Delta\langle \hat{A}\rangle(t,|\Psi(0)\rangle )$ and $\Delta\langle \hat{A}\rangle(t,|\Psi_{r}(0)\rangle ) $ with respect to time $t$ is not zero for the first time. Here, we take $\left|\Psi_{r}(0)\right\rangle$ as a known initial state, which is called the reference state, and $\left|\Psi(0)\right\rangle$ to be an unknown state. Therefore, the average value of a certain Hermitian operator $[i\hat{H},\cdots[i\hat{H},[i\hat{H},\hat{A}]]]$ in the initial state $\left|\Psi(0)\right\rangle$ can be obtained by measuring the variation of the Hermitian operator $\hat{A}$ corresponding to the observable in a short time, thus we can obtain the information about the unknown state $\left|\Psi(0)\right\rangle$, i.e.,
\begin{equation}
\left\langle \Psi(0)\right|[\underset{n}{\underbrace{i\hat{H},\cdots[i\hat{H},[i\hat{H}}},\hat{A}]]]\left|\Psi(0)\right\rangle=\left\langle \Psi_{r}(0)\right|[\underset{n}{\underbrace{i\hat{H},\cdots[i\hat{H},[i\hat{H}}},\hat{A}]]]\left|\Psi_{r}(0)\right\rangle \lim_{t\rightarrow0}\frac{\Delta\langle \hat{A}\rangle(t,|\Psi(0)\rangle ) }{\Delta\langle \hat{A}\rangle(t,|\Psi_{r}(0)\rangle ) }.  \tag{S8}\label{S8}
\end{equation} 
When the initial state of the system is a mixed state, its initial density matrix operator is $\hat{\rho}(0)$. Then, the density matrix of the system at time $t$ is
\begin{equation}
\hat{\rho}(t)=e^{-i\hat{H}t}\hat{\rho}(0)e^{i\hat{H}t}. \tag{S9}\label{S9}
\end{equation}
The $n$th order derivative of the change in the mean value of the operator $\hat{A}$ at time $t$ with respect to time $t$ is as follows
\allowdisplaybreaks
\begin{align}
\frac{d^{n}\Delta\langle \hat{A}\rangle (t,\hat{\rho}(0))}{dt^{n}}&=\frac{d^{n}}{dt^{n}}\Big[{\rm Tr}[\hat{\rho}(t)\hat{A}]-{\rm Tr}[\hat{\rho}(0)\hat{A}]\Big] \nonumber \\
&={\rm Tr}\left[\frac{d^{n}\hat{\rho}(t)}{dt^{n}}\hat{A}\right] \nonumber \\
&={\rm Tr}\left[[\underset{n}{\underbrace{-i\hat{H},\cdots[-i\hat{H},[-i\hat{H}}},\hat{\rho}(0)]]]\hat{A}\right] \nonumber \\
&={\rm Tr}\left[[-i\hat{H},[\underset{n-1}{\underbrace{-i\hat{H},\cdots[-i\hat{H},[-i\hat{H}}},\hat{\rho}(0)]]]]\hat{A}\right] \nonumber \\
&={\rm Tr}\left[[\underset{n-1}{\underbrace{-i\hat{H},\cdots[-i\hat{H},[-i\hat{H}}},\hat{\rho}(0)]]][i\hat{H},\hat{A}]\right]  \nonumber \\
&={\rm Tr}\left[[-i\hat{H},[\underset{n-2}{\underbrace{-i\hat{H},\cdots[-i\hat{H},[-i\hat{H}}},\hat{\rho}(0)]]]][i\hat{H},\hat{A}]\right] \nonumber \\
&={\rm Tr}\left[[\underset{n-2}{\underbrace{-i\hat{H},\cdots[-i\hat{H},[-i\hat{H}}},\hat{\rho}(0)]]][i\hat{H},[i\hat{H},\hat{A}]]\right] \nonumber \\
&={\rm Tr}\left[[-i\hat{H},\hat{\rho}(0)][\underset{n-1}{\underbrace{i\hat{H},\cdots[i\hat{H},[i\hat{H}}},\hat{A}]]]\right] \nonumber \\
&={\rm Tr}\left[[[\underset{n}{\underbrace{i\hat{H},\cdots[i\hat{H},[i\hat{H}}},\hat{A}]]]\hat{\rho}(0)]\right].\tag{S10}\label{S10}
\end{align}
Similar to the above, when the system starts evolving with another known reference mixed state $\hat{\rho}_{r}(0)$, we can obtain
\begin{equation}
{\rm Tr}\left[[[\underset{n}{\underbrace{i\hat{H},\cdots[i\hat{H},[i\hat{H}}},\hat{A}]]]\hat{\rho}(0)]\right]={\rm Tr}\left[[[\underset{n}{\underbrace{i\hat{H},\cdots[i\hat{H},[i\hat{H}}},\hat{A}]]]\hat{\rho}_{r}(0)]\right]\lim_{t\rightarrow0}\frac{\Delta\langle \hat{A}\rangle (t,\hat{\rho}(0))}{\Delta\langle \hat{A}\rangle (t,\hat{\rho}_{r}(0))}.\tag{S11}\label{S11}
\end{equation}
When both $\hat{\rho}(0)$ and $\hat{\rho}_{r}(0)$ are pure states, the above equation is the same as Eq.~(\ref{S8}).
This method of measuring the instantaneous change of a mechanical quantity to obtain information about another mechanical quantity in a certain initial state is called the instantaneous indirect measurement method. In particular, when $\langle\hat{A}\rangle$ and $\langle[i\hat{H},\cdots[i\hat{H},[i\hat{H},\hat{A}]]]\rangle$ belong to different subsystems, we can get the information of one subsystem just by measuring the change of the operator of the another subsystem, and this measurement can avoid the influence of the measurement on the target system as much as possible. We will emphasize this point later in the specific model.

\subsection{Applications of the IIMP}
In the following, we will discuss the application of IIMP to some typical models, such as the $p$-photon Rabi model, the $p$-photon Jaynes-Cummings (JC) model, the $p$-photon Dicke model, and the $p$-photon Tavis-Cummings (TC) model. First, we consider a extended $p$-photon Rabi model with the following Hamiltonian
\begin{equation}
\hat{H}_{R}=\omega_{a}\text{\ensuremath{\hat{a}^{\dagger}}}\hat{a}+\frac{\omega_{0}}{2}\hat{\sigma}_{z}+g(\hat{a}^{\dagger p}+\hat{a}^{p})(\hat{\sigma}_{-}+\hat{\sigma}_{+})+\frac{U}{2}\hat{a}^{\dagger2}\hat{a}^{2}+\gamma\text{\ensuremath{\hat{a}^{\dagger}}}\hat{a}\hat{\sigma}_{z}, \tag{S12}\label{S12}
\end{equation}
where $p$ is an integer greater than or equal to $1$. $\omega_{a}$ and $\omega_{0}$ are the eigenfrequency of the optical cavity and the transition frequency of the two-level atom, respectively. $g$ is the dipole interaction strength between the cavity field and the single atom, $U$ is the Kerr-type nonlinear interaction strength of the optical field, and $\gamma$ is the dispersion interaction between the cavity field and the single atom. $\hat{a}$ and $\hat{a}^{\dagger}$ are, respectively, the annihilation and creation operators of the single-mode cavity field, they satisfy the usual bosonic commutation relation $[\hat{a},\hat{a}^{\dagger}]=1$. $\hat{\sigma}_{x,y,z}$ are the usual Pauli operators, and $\hat{\sigma}_{\pm}=\frac{1}{2}(\hat{\sigma}_{x}\pm i\hat{\sigma}_{y})$. Hereafter we set $\hbar=1$.

In the regime of near resonance $\omega_{a}\approx\omega_{a}$ and relatively weak coupling $g\ll \min\{\omega_{0},\omega_{a}\}$, we can obtain the Hamiltonian of the $p$-photon JC model under the rotating-wave approximation as
\begin{equation}
\hat{H}_{JC}=\omega_{a}\text{\ensuremath{\hat{a}^{\dagger}}}\hat{a}+\frac{\omega_{0}}{2}\hat{\sigma}_{z}+g(\hat{a}^{\dagger p}\hat{\sigma}_{-}+\hat{\sigma}_{+}\hat{a}^{p})+\frac{U}{2}\hat{a}^{\dagger2}\hat{a}^{2}+\gamma\text{\ensuremath{\hat{a}^{\dagger}}}\hat{a}\hat{\sigma}_{z}.  \tag{S13}\label{S13}
\end{equation}

\begin{figure}[t]
	\setlength{\abovecaptionskip}{0.cm}
	\setlength{\belowcaptionskip}{-0.cm}
	\centering
	\includegraphics[width=8cm,height=6cm]{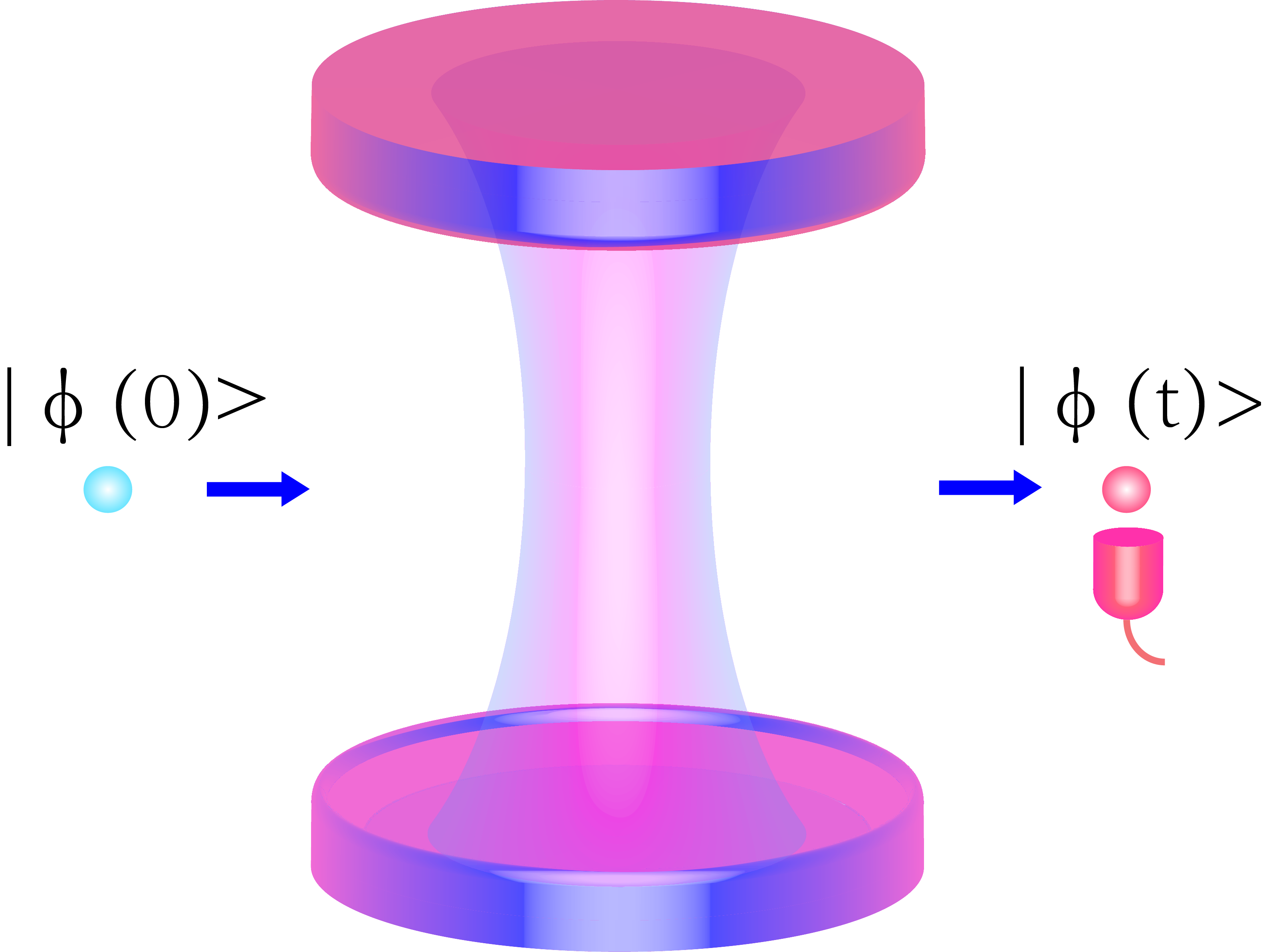}
	\caption{A schematic diagram for indirectly obtaining information about the cavity field by measuring the change of the atomic energy. According to the measurement target, the initial state of the atom is prepared in the known state $\left|\phi(0)\right\rangle$. The state of the atom after instantaneous interaction with the cavity field is $\left|\phi(t)\right\rangle$. The information of the cavity field can be obtained by measuring the energy change of the atom. This measurement does not have additional effects on the cavity field except for the instantaneous interaction.}
	\label{figS1}
\end{figure}

First, we take an unknown state of the cavity field as the object we need to study. Here, we want to obtain information about this unknown state of the cavity field by measuring the change of atomic energy. Therefore, we choose the energy of the atom as the observation quantity to be measured. For the Rabi and JC models, when the initial states of the cavity field are $|\psi(0)\rangle$ and $|\psi_{r}(0)\rangle$, respectively, we can construct the equations separately as follows
\begin{equation}
\lim_{t\rightarrow0}\frac{\Delta\left\langle \hat{\sigma}_{z}\right\rangle(t,\left|\psi(0)\right\rangle ) }{\Delta\left\langle \hat{\sigma}_{z}\right\rangle(t,\left|\psi_{r}(0)\right\rangle ) }=\frac{\left\langle \psi(0)\right|[\underset{n}{\underbrace{i\hat{H}_{R},\cdots[i\hat{H}_{R},[i\hat{H}_{R}}},\hat{\sigma}_{z}]]]\left|\psi(0)\right\rangle }{\left\langle \psi_{r}(0)\right|[\underset{n}{\underbrace{i\hat{H}_{R},\cdots[i\hat{H}_{R},[i\hat{H}_{R}}},\hat{\sigma}_{z}]]]\left|\psi_{r}(0)\right\rangle }, \tag{S14}\label{S14}
\end{equation}
and
\begin{equation}
\lim_{t\rightarrow0}\frac{\Delta\left\langle \hat{\sigma}_{z}\right\rangle(t,\left|\psi(0)\right\rangle ) }{\Delta\left\langle \hat{\sigma}_{z}\right\rangle(t,\left|\psi_{r}(0)\right\rangle ) }=\frac{\left\langle \psi(0)\right|[\underset{n}{\underbrace{i\hat{H}_{JC},\cdots[i\hat{H}_{JC},[i\hat{H}_{JC}}},\hat{\sigma}_{z}]]]\left|\psi(0)\right\rangle }{\left\langle \psi_{r}(0)\right|[\underset{n}{\underbrace{i\hat{H}_{JC},\cdots[i\hat{H}_{JC},[i\hat{H}_{JC}}},\hat{\sigma}_{z}]]]\left|\psi_{r}(0)\right\rangle }, \tag{S15}\label{S15}
\end{equation}
where $n$ is determined by when the $n$th-order derivative of $\Delta\left\langle \hat{\sigma}_{z}\right\rangle(t,\left|\psi(0)\right\rangle )$ and $\Delta\left\langle \hat{\sigma}_{z}\right\rangle(t,\left|\psi_{r}(0)\right\rangle )$ with respect to time $t$ is not zero for the first time, and $\left|\psi_{r}(0)\right\rangle$ is a known reference initial state. $\left|\psi(0)\right\rangle$ is either a completely unknown initial state or the initial state of only some of these subsystems is unknown, depending on the specific object of study.
In the following, we start to analyze the first-order derivatives of 
$\Delta\left\langle \hat{\sigma}_{z}\right\rangle(t,\left|\psi(0)\right\rangle )$ and $\Delta\left\langle \hat{\sigma}_{z}\right\rangle(t,\left|\psi_{r}(0)\right\rangle )$ with respect to time $t$ when $t\rightarrow0$. In the Rabi model and JC model, these two derivatives are respectively
\begin{align}
	\lim_{t\rightarrow0}\frac{d\Delta\left\langle \hat{\sigma}_{z}\right\rangle(t,\left|\psi(0)\right\rangle )}{dt}&=\left[i\hat{H}_{R},\hat{\sigma}_{z}\right]\nonumber\\
	&=\left[ig(\hat{a}^{\dagger p}+\hat{a}^{p})(\hat{\sigma}_{-}+\hat{\sigma}_{+}),\hat{\sigma}_{z}\right]    \nonumber \\ 
	&=2ig(\hat{a}^{\dagger p}+\hat{a}^{p})(\hat{\sigma}_{-}-\hat{\sigma}_{+}),   \tag{S16}\label{S16}
\end{align}
and
\begin{align}
	\lim_{t\rightarrow0}\frac{d\Delta\left\langle \hat{\sigma}_{z}\right\rangle(t,\left|\psi(0)\right\rangle )}{dt}&=\left[i\hat{H}_{JC},\hat{\sigma}_{z}\right] \nonumber\\
	&=\left[ig(\hat{a}^{\dagger p}\hat{\sigma}_{-}+\hat{\sigma}_{+}\hat{a}^{p}),\hat{\sigma}_{z}\right]  \nonumber \\
	&=2ig\left(\hat{a}^{\dagger p}\hat{\sigma}_{-}-\hat{\sigma}_{+}\hat{a}^{p}\right).\tag{S17}\label{S17}
\end{align}
Obviously, when an arbitrary initial state $\left|\psi_{a}(0)\right\rangle$ and the reference state  $\left|\psi_{r}(0)\right\rangle$ of the Rabi model satisfy respectively the constraint $\left\langle\psi_{a}(0)\right|(\hat{a}^{\dagger p}+\hat{a}^{p})(\hat{\sigma}_{-}-\hat{\sigma}_{+})\left|\psi_{a}(0)\right\rangle\neq0$ and $\left\langle\psi_{r}(0)\right|(\hat{a}^{\dagger p}+\hat{a}^{p})(\hat{\sigma}_{-}-\hat{\sigma}_{+})\left|\psi_{r}(0)\right\rangle\neq0$, then we can obtain the average value of the Hermitian operator $i(\hat{a}^{\dagger p}+\hat{a}^{p})(\hat{\sigma}_{-}-\hat{\sigma}_{+})$ in this arbitrary initial state by measuring the change of the atomic energy, thus achieving an indirect measure of the Hermitian operator $i(\hat{a}^{\dagger p}+\hat{a}^{p})(\hat{\sigma}_{-}-\hat{\sigma}_{+})$, i.e.,
\begin{equation}
\left\langle\psi_{a}(0)\right|i(\hat{a}^{\dagger p}+\hat{a}^{p})(\hat{\sigma}_{-}-\hat{\sigma}_{+})\left|\psi_{a}(0)\right\rangle=\lim_{t\rightarrow0}\frac{\Delta\left\langle \hat{\sigma}_{z}\right\rangle(t,\left|\psi_{a}(0)\right\rangle ) }{\Delta\left\langle \hat{\sigma}_{z}\right\rangle(t,\left|\psi_{r}(0)\right\rangle ) }\left\langle\psi_{r}(0)\right|i(\hat{a}^{\dagger p}+\hat{a}^{p})(\hat{\sigma}_{-}-\hat{\sigma}_{+})\left|\psi_{r}(0)\right\rangle. \tag{S18}\label{S18}
\end{equation}
Similarly, for the JC model, if the initial state $\left|\psi_{a}(0)\right\rangle$ and the reference state  $\left|\psi_{r}(0)\right\rangle$ satisfy the constraint $\left\langle\psi_{a}(0)\right|i\left(\hat{a}^{\dagger p}\hat{\sigma}_{-}-\hat{\sigma}_{+}\hat{a}^{p}\right)\left|\psi_{a}(0)\right\rangle\neq0$ and $\left\langle\psi_{r}(0)\right|i\left(\hat{a}^{\dagger p}\hat{\sigma}_{-}-\hat{\sigma}_{+}\hat{a}^{p}\right)\left|\psi_{r}(0)\right\rangle\neq0$, respectively, then we can similarly obtain the mean value of the Hermitian operator $i\left(\hat{a}^{\dagger p}\hat{\sigma}_{-}-\hat{\sigma}_{+}\hat{a}^{p}\right)$ in the initial state $\left|\psi_{a}(0)\right\rangle$ by measuring the change of the atomic energy 
\begin{equation}
\left\langle\psi_{a}(0)\right|i\left(\hat{a}^{\dagger p}\hat{\sigma}_{-}-\hat{\sigma}_{+}\hat{a}^{p}\right)\left|\psi_{a}(0)\right\rangle=\left\langle\psi_{r}(0)\right|i\left(\hat{a}^{\dagger p}\hat{\sigma}_{-}-\hat{\sigma}_{+}\hat{a}^{p}\right)\left|\psi_{r}(0)\right\rangle\lim_{t\rightarrow0}\frac{\Delta\left\langle \hat{\sigma}_{z}\right\rangle(t,\left|\psi_{a}(0)\right\rangle ) }{\Delta\left\langle \hat{\sigma}_{z}\right\rangle(t,\left|\psi_{r}(0)\right\rangle ) }.  \tag{S19}\label{S19}
\end{equation}
Obviously, when $p=1$ and the initial state of the total system is prepared in $|\psi_{a}(0)\rangle=|\varphi_{a}\rangle\otimes|\phi\rangle=|\varphi_{a}\rangle\otimes(|g\rangle-ie^{i\theta}|e\rangle)/\sqrt{2}$, i.e., the cavity field is in an arbitrary state $|\varphi_{a}\rangle$ and the atom is in state $|\phi\rangle=(|g\rangle-ie^{i\theta}|e\rangle)/\sqrt{2}$, we can measure the quadrature operators  $\hat{X}(\theta)$ by measuring the change in the energy of the probe atom. This indirect measurement method has almost no effect on the optical field to be measured compared to balanced homodyne detection.

However, when the initial state of the Rabi model or JC model is $\left|\psi(0)\right\rangle=\left|\varphi_{a}(0)\right\rangle\otimes\left|g\right\rangle$ (or $\otimes \left|e\right\rangle$), i.e., the cavity field is in an arbitrary state $\left|\varphi_{a}(0)\right\rangle$ and the atom is in the ground state $\left|g\right\rangle$ (or excited state $\left|e\right\rangle$), then
\begin{align}
\lim_{t\rightarrow0}\frac{d\Delta\left\langle \hat{\sigma}_{z}\right\rangle(t,\left|\psi(0)\right\rangle) }{dt}&=\left\langle \psi(0)\right|[i\hat{H}_{R},\hat{\sigma}_{z}]\left|\psi(0)\right\rangle =0,  \tag{S20}\label{S20}\\
\lim_{t\rightarrow0}\frac{d\Delta\left\langle \hat{\sigma}_{z}\right\rangle(t,\left|\psi(0)\right\rangle) }{dt}&=\left\langle \psi(0)\right|[i\hat{H}_{JC},\hat{\sigma}_{z}]\left|\psi(0)\right\rangle =0 . \tag{S21}\label{S21}
\end{align}
In this case, we need to consider the second-order derivative of $\Delta\left\langle \hat{\sigma}_{z}\right\rangle(t,\left|\psi(0)\right\rangle)$  with respect to $t$. In the Rabi model and JC model, these two second-order derivatives are respectively
\begin{align}
	\lim_{t\rightarrow0}\frac{d^{2}\Delta\left\langle \hat{\sigma}_{z}\right\rangle(t,\left|\psi(0)\right\rangle) }{dt^{2}}&=\left\langle \psi(0)\right|\left[i\hat{H}_{Rabi},\left[i\hat{H}_{Rabi},\hat{A}\right]\left|\psi(0)\right\rangle\right]\nonumber\\
	&=\left\langle \psi(0)\right|\left[i\hat{H}_{Rabi},2ig(\hat{a}^{\dagger p}+\hat{a}^{p})(\hat{\sigma}_{-}-\hat{\sigma}_{+})\right]\left|\psi(0)\right\rangle\nonumber \\&=-2g\left\langle \psi(0)\right|\left(\begin{array}{c}
		\left[\omega_{a}\text{\ensuremath{\hat{a}^{\dagger}}}\hat{a}+\frac{\omega_{0}}{2}\hat{\sigma}_{z}+\frac{U}{2}\hat{a}^{\dagger2}\hat{a}^{2}+\gamma\text{\ensuremath{\hat{a}^{\dagger}}}\hat{a}\hat{\sigma}_{z},(\hat{a}^{\dagger p}+\hat{a}^{p})(\hat{\sigma}_{-}-\hat{\sigma}_{+})\right]\nonumber\\
		+\left[g(\hat{a}^{\dagger p}+\hat{a}^{p})(\hat{\sigma}_{-}+\hat{\sigma}_{+}),(\hat{a}^{\dagger p}+\hat{a}^{p})(\hat{\sigma}_{-}-\hat{\sigma}_{+})\right]
	\end{array}\right)\left|\psi(0)\right\rangle\nonumber\\&=-2g\left\langle \psi(0)\right|\left(\begin{array}{c}
		\left[\omega_{a}\text{\ensuremath{\hat{a}^{\dagger}}}\hat{a}+\frac{\omega_{0}}{2}\hat{\sigma}_{z}+\frac{U}{2}\hat{a}^{\dagger2}\hat{a}^{2}+\gamma\text{\ensuremath{\hat{a}^{\dagger}}}\hat{a}\hat{\sigma}_{z},(\hat{a}^{\dagger p}+\hat{a}^{p})(\hat{\sigma}_{-}-\hat{\sigma}_{+})\right]\\
		+2g(\hat{a}^{\dagger p}+\hat{a}^{p})^{2}\hat{\sigma}_{z}
	\end{array}\right)\left|\psi(0)\right\rangle, \tag{S22}\label{S22}\\
	\lim_{t\rightarrow0}\frac{d^{2}\Delta\left\langle \hat{\sigma}_{z}\right\rangle(t,\left|\psi(0)\right\rangle) }{dt^{2}}&=\left\langle \psi(0)\right|\left[i\hat{H}_{JC},\left[i\hat{H}_{JC},\hat{A}\right]\right]\left|\psi(0)\right\rangle\nonumber\\
	&=\left\langle \psi(0)\right|\left[i\hat{H}_{JC},2ig\left(\hat{a}^{\dagger p}\hat{\sigma}_{-}-\hat{\sigma}_{+}\hat{a}^{p}\right)\right]\left|\psi(0)\right\rangle\nonumber\\&=-2g\left\langle \psi(0)\right|\left(\begin{array}{c}
		\left[\omega_{a}\text{\ensuremath{\hat{a}^{\dagger}}}\hat{a}+\frac{\omega_{0}}{2}\hat{\sigma}_{z}+\frac{U}{2}\hat{a}^{\dagger2}\hat{a}^{2}+\gamma\text{\ensuremath{\hat{a}^{\dagger}}}\hat{a}\hat{\sigma}_{z},\hat{a}^{\dagger p}\hat{\sigma}_{-}-\hat{\sigma}_{+}\hat{a}^{p}\right]\nonumber\\
		+\left[g(\hat{a}^{\dagger p}\hat{\sigma}_{-}+\hat{\sigma}_{+}\hat{a}^{p}),\hat{a}^{\dagger p}\hat{\sigma}_{-}-\hat{\sigma}_{+}\hat{a}^{p}\right]
	\end{array}\right)\left|\psi(0)\right\rangle \nonumber\\&=-2g\left\langle \psi(0)\right|\left(\begin{array}{c}
		\left[\omega_{a}\text{\ensuremath{\hat{a}^{\dagger}}}\hat{a}+\frac{\omega_{0}}{2}\hat{\sigma}_{z}+\frac{U}{2}\hat{a}^{\dagger2}\hat{a}^{2}+\gamma\text{\ensuremath{\hat{a}^{\dagger}}}\hat{a}\hat{\sigma}_{z},\hat{a}^{\dagger p}\hat{\sigma}_{-}-\hat{\sigma}_{+}\hat{a}^{p}\right]\\
		+2g\hat{a}^{p}\hat{a}^{\dagger p}\hat{\sigma}_{+}\hat{\sigma}_{-}-2g\hat{a}^{\dagger p}\hat{a}^{p}\hat{\sigma}_{-}\hat{\sigma}_{+}
	\end{array}\right)\left|\psi(0)\right\rangle   . \tag{S23}\label{S23}
\end{align}
Thus, for the Rabi model, when we prepare the initial state of the atom in the ground $\left|g\right\rangle$ or excited state $\left|e\right\rangle$, we are able to obtain the average value of the Hermitian operator $(\hat{a}^{\dagger p}+\hat{a}^{p})^{2}$ in the initial state $\left|\varphi_{a}(0)\right\rangle$ of the optical field by the variation of the average value of $\hat{\sigma}_{z}$, i.e.,
\begin{equation}
\left\langle\varphi_{a}(0)\right|(\hat{a}^{\dagger p}+\hat{a}^{p})^{2}\left|\varphi_{a}(0)\right\rangle=\left\langle\varphi_{r}(0)\right|(\hat{a}^{\dagger p}+\hat{a}^{p})^{2}\left|\varphi_{r}(0)\right\rangle\lim_{t\rightarrow0}\frac{\Delta\left\langle \hat{\sigma}_{z}\right\rangle(t,\left|\psi(0)\right\rangle ) }{\Delta\left\langle \hat{\sigma}_{z}\right\rangle(t,\left|\psi_{r}(0)\right\rangle ) } ,  \tag{S24}\label{S24}
\end{equation}
where this known reference state is $\left|\psi_{r}(0)\right\rangle=\left|\varphi_{r}(0)\right\rangle\otimes\left|g\right\rangle$ (or $\otimes \left|e\right\rangle$), and $\left|\varphi_{r}(0)\right\rangle$ is a known optical field state. For example, we can choose it as a number state $\left|\varphi_{r}(0)\right\rangle=\left|1\right\rangle$. 
For the JC model, when the cavity field is in an arbitrary state that is not a vacuum, if the atom is initially in the ground state $\left|g\right\rangle$, we can obtain the $p$-order correlation function $\left\langle\hat{a}^{\dagger p}\hat{a}^{p}\right\rangle$ of the cavity field by measuring the change in the mean value of $\hat{\sigma}_{z}$, i.e.,
\begin{equation}
\left\langle\varphi_{a}(0)\right|\hat{a}^{\dagger p}\hat{a}^{p}\left|\varphi_{a}(0)\right\rangle=\left\langle\varphi_{r}(0)\right|\hat{a}^{\dagger p}\hat{a}^{p}\left|\varphi_{r}(0)\right\rangle\lim_{t\rightarrow0}\frac{\Delta\left\langle \hat{\sigma}_{z}\right\rangle(\left|\psi(0)\right\rangle ) }{\Delta\left\langle \hat{\sigma}_{z}\right\rangle(\left|\psi_{r}(0)\right\rangle ) }. \tag{S25}\label{S25}
\end{equation}
It is worth noting that here we need to select the reference state $\left|\varphi_{r}(0)\right\rangle$ according to the size of $p$ to prevent the appearance of $\left\langle\varphi_{r}(0)\right|\hat{a}^{\dagger p}\hat{a}^{p}\left|\varphi_{r}(0)\right\rangle=0$. In contrast, if the atom is initially in an excited state $\left|e\right\rangle$, we can obtain the average value of the Hermitian operator $\hat{a}^{p}\hat{a}^{\dagger p}$ in the initial state $\left|\varphi_{a}(0)\right\rangle$ by measuring the change in the average value of $\hat{\sigma}_{z}$, i.e.,
\begin{equation}
\left\langle\varphi_{a}(0)\right|\hat{a}^{p}\hat{a}^{\dagger p}\left|\varphi_{a}(0)\right\rangle=\left\langle\varphi_{r}(0)\right|\hat{a}^{p}\hat{a}^{\dagger p}\left|\varphi_{r}(0)\right\rangle\lim_{t\rightarrow0}\frac{\Delta\left\langle \hat{\sigma}_{z}\right\rangle(t,\left|\psi(0)\right\rangle ) }{\Delta\left\langle \hat{\sigma}_{z}\right\rangle(t,\left|\psi_{r}(0)\right\rangle ) }. \tag{S26}\label{S26}
\end{equation}

Second, we take an unknown state of the atom as the object of study, and we hope to obtain information about this unknown state of the atom by measuring the average photon number of the cavity field. Here, we choose the average photon number of the cavity field as the measurement object. 
Similar to above, for the Rabi and JC model, we can construct two equations as follows
\begin{equation}
\lim_{t\rightarrow0}\frac{\Delta\left\langle \hat{a}^{\dagger}\hat{a}\right\rangle(t,\left|\psi(0)\right\rangle ) }{\Delta\left\langle \hat{a}^{\dagger}\hat{a}\right\rangle(t,\left|\psi_{r}(0)\right\rangle ) }=\frac{\left\langle \psi(0)\right|[\underset{n}{\underbrace{i\hat{H}_{R},\cdots[i\hat{H}_{R},[i\hat{H}_{R}}},\hat{a}^{\dagger}\hat{a}]]]\left|\psi(0)\right\rangle }{\left\langle \psi_{r}(0)\right|[\underset{n}{\underbrace{i\hat{H}_{R},\cdots[i\hat{H}_{R},[i\hat{H}_{R}}},\hat{a}^{\dagger}\hat{a}]]]\left|\psi_{r}(0)\right\rangle } ,  \tag{S27}\label{S27}
\end{equation}
and
\begin{equation}
\lim_{t\rightarrow0}\frac{\Delta\left\langle \hat{a}^{\dagger}\hat{a}\right\rangle(t,\left|\psi(0)\right\rangle ) }{\Delta\left\langle \hat{a}^{\dagger}\hat{a}\right\rangle(t,\left|\psi_{r}(0)\right\rangle ) }=\frac{\left\langle \psi(0)\right|[\underset{n}{\underbrace{i\hat{H}_{JC},\cdots[i\hat{H}_{JC},[i\hat{H}_{JC}}},\hat{a}^{\dagger}\hat{a}]]]\left|\psi(0)\right\rangle }{\left\langle \psi_{r}(0)\right|[\underset{n}{\underbrace{i\hat{H}_{JC},\cdots[i\hat{H}_{JC},[i\hat{H}_{JC}}},\hat{a}^{\dagger}\hat{a}]]]\left|\psi_{r}(0)\right\rangle } .  \tag{S28}\label{S28}
\end{equation}
In the Rabi and JC models, the first-order derivatives of $\Delta\left\langle \hat{a}^{\dagger}\hat{a}\right\rangle(t,\left|\psi(0)\right\rangle ) $ with respect to time $t$ are respectively as follows
\begin{align}
	\lim_{t\rightarrow0}\frac{d\Delta\left\langle \hat{a}^{\dagger}\hat{a}\right\rangle (t,\left|\psi(0)\right\rangle )}{dt }&=\left\langle \psi(0)\right|\left[i\hat{H}_{R},\hat{a}^{\dagger}\hat{a}\right]\left|\psi(0)\right\rangle\nonumber\\
	&=\left\langle \psi(0)\right|\left[ig(\hat{a}^{\dagger p}+\hat{a}^{p})(\hat{\sigma}_{-}+\hat{\sigma}_{+}),\text{\ensuremath{\hat{a}^{\dagger}}}\hat{a}\right]\left|\psi(0)\right\rangle \nonumber\\
	&=ig\left\langle \psi(0)\right|[(\hat{\sigma}_{-}+\hat{\sigma}_{+})\left(\hat{a}^{\dagger p}\text{\ensuremath{\hat{a}^{\dagger}}}\hat{a}+\hat{a}^{p}\text{\ensuremath{\hat{a}^{\dagger}}}\hat{a}-\text{\ensuremath{\hat{a}^{\dagger}}}\hat{a}\hat{a}^{\dagger p}-\text{\ensuremath{\hat{a}^{\dagger}}}\hat{a}\hat{a}^{p}\right)]\left|\psi(0)\right\rangle \nonumber \\
	&=ipg\left\langle \psi(0)\right|(\hat{\sigma}_{-}+\hat{\sigma}_{+})\left(\hat{a}^{p}-\hat{a}^{\dagger p}\right)\left|\psi(0)\right\rangle ,\tag{S29}\label{S29}\\
	\lim_{t\rightarrow0}\frac{d\Delta\left\langle \hat{a}^{\dagger}\hat{a}\right\rangle(t,\left|\psi(0)\right\rangle ) }{dt }&=\left\langle \psi(0)\right|\left[i\hat{H}_{JC},\hat{a}^{\dagger}\hat{a}\right]\left|\psi(0)\right\rangle \nonumber\\
	&=\left\langle \psi(0)\right|\left[ig(\hat{a}^{\dagger p}\hat{\sigma}_{-}+\hat{\sigma}_{+}\hat{a}^{p}),\hat{a}^{\dagger}\hat{a}\right]\left|\psi(0)\right\rangle\nonumber\\
	&=ig\left\langle \psi(0)\right|\left(\hat{a}^{\dagger p}\hat{a}^{\dagger}\hat{a}\hat{\sigma}_{-}+\hat{\sigma}_{+}\hat{a}^{p}\hat{a}^{\dagger}\hat{a}-\hat{a}^{\dagger}\hat{a}\hat{a}^{\dagger p}\hat{\sigma}_{-}-\hat{\sigma}_{+}\hat{a}^{\dagger}\hat{a}\hat{a}^{p}\right)\left|\psi(0)\right\rangle\nonumber\\
	&=ig\left\langle \psi(0)\right|\left(\left(\hat{a}^{\dagger}\hat{a}-p\right)\hat{a}^{\dagger p}\hat{\sigma}_{-}+\hat{\sigma}_{+}\left(\hat{a}^{\dagger}\hat{a}+p\right)\hat{a}^{p}-\hat{a}^{\dagger}\hat{a}\hat{a}^{\dagger p}\hat{\sigma}_{-}-\hat{\sigma}_{+}\hat{a}^{\dagger}\hat{a}\hat{a}^{p}\right)\left|\psi(0)\right\rangle\nonumber\\
	&=ipg\left\langle \psi(0)\right|\left(\hat{\sigma}_{+}\hat{a}^{p}-\hat{a}^{\dagger p}\hat{\sigma}_{-}\right)\left|\psi(0)\right\rangle . \tag{S30}\label{S30}
\end{align}
When $p = 1$ and the initial state of the cavity field is a vacuum or coherent state, we have studied the examples in the main text and will not discuss them here.

When the initial state of the Rabi model or JC model is $\left|\psi(0)\right\rangle=\left|N\right\rangle\otimes\left|\phi_{a}\right\rangle$, i.e.,
the cavity field is prepared in an arbitrary number state $|\varphi(0)\rangle=|N\rangle$ and the atom is in an arbitrary state $\left|\phi_{a}\right\rangle$, we can obtain
\begin{align}
	\lim_{t\rightarrow0}\frac{d\Delta\left\langle \hat{a}^{\dagger}\hat{a}\right\rangle(t,\left|\psi(0)\right\rangle ) }{dt }&=\left\langle \psi(0)\right|\left[i\hat{H}_{R},\hat{a}^{\dagger}\hat{a}\right]\left|\psi(0)\right\rangle =0, \tag{S31}\label{S31}\\
	\lim_{t\rightarrow0}\frac{d\Delta\left\langle \hat{a}^{\dagger}\hat{a}\right\rangle(t,\left|\psi(0)\right\rangle ) }{dt }&=\left\langle \psi(0)\right|\left[i\hat{H}_{JC},\hat{a}^{\dagger}\hat{a}\right]\left|\psi(0)\right\rangle =0 . \tag{S32}\label{S32}
\end{align}
Then we need to consider the second order derivative of $\Delta\left\langle \hat{a}^{\dagger}\hat{a}\right\rangle(t,\left|\psi(0)\right\rangle )$  with respect to $t$. In the Rabi and JC models, these two second-order derivatives are respectively as follows
\begin{align}
	\lim_{t\rightarrow0}\frac{d^{2}\Delta\left\langle \hat{a}^{\dagger}\hat{a}\right\rangle(t,\left|\psi(0)\right\rangle ) }{dt^{2} }&=\left\langle \psi(0)\right|\left[i\hat{H}_{R},\left[i\hat{H}_{R},\hat{a}^{\dagger}\hat{a}\right]\right]\left|\psi(0)\right\rangle\nonumber\\
	&=\left\langle \psi(0)\right|\left[i\hat{H}_{R},ipg(\hat{\sigma}_{-}+\hat{\sigma}_{+})\left(\hat{a}^{p}-\hat{a}^{\dagger p}\right)\right]\left|\psi(0)\right\rangle\nonumber\\
	&=-pg\left\langle \psi(0)\right|\left(\begin{array}{c}
		\left[\omega_{a}\text{\ensuremath{\hat{a}^{\dagger}}}\hat{a}+\frac{\omega_{0}}{2}\hat{\sigma}_{z}+\frac{U}{2}\hat{a}^{\dagger2}\hat{a}^{2}+\gamma\text{\ensuremath{\hat{a}^{\dagger}}}\hat{a}\hat{\sigma}_{z},(\hat{\sigma}_{-}+\hat{\sigma}_{+})\left(\hat{a}^{p}-\hat{a}^{\dagger p}\right)\right]\\
		+\left[g(\hat{a}^{\dagger p}+\hat{a}^{p})(\hat{\sigma}_{-}+\hat{\sigma}_{+}),(\hat{\sigma}_{-}+\hat{\sigma}_{+})\left(\hat{a}^{p}-\hat{a}^{\dagger p}\right)\right]
	\end{array}\right)\left|\psi(0)\right\rangle \nonumber\\
	&=-pg\left\langle \psi(0)\right|\left(\begin{array}{c}
		\left[\omega_{a}\text{\ensuremath{\hat{a}^{\dagger}}}\hat{a}+\frac{\omega_{0}}{2}\hat{\sigma}_{z}+\frac{U}{2}\hat{a}^{\dagger2}\hat{a}^{2}+\gamma\text{\ensuremath{\hat{a}^{\dagger}}}\hat{a}\hat{\sigma}_{z},(\hat{\sigma}_{-}+\hat{\sigma}_{+})\left(\hat{a}^{p}-\hat{a}^{\dagger p}\right)\right]\\
		+2g\left(\hat{a}^{\dagger p}\hat{a}^{p}-\hat{a}^{p}\hat{a}^{\dagger p}\right)
	\end{array}\right)\left|\psi(0)\right\rangle,  \tag{S33}\label{S33}\\
	\lim_{t\rightarrow0}\frac{d^{2}\Delta\left\langle \hat{a}^{\dagger}\hat{a}\right\rangle(t,\left|\psi(0)\right\rangle ) }{dt^{2} }&=\left\langle \psi(0)\right|\left[i\hat{H}_{JC},\left[i\hat{H}_{JC},\hat{a}^{\dagger}\hat{a}\right]\right]\left|\psi(0)\right\rangle\nonumber\\
	&=\left\langle \psi(0)\right|\left[i\hat{H}_{JC},-ipg\left(\hat{a}^{\dagger p}\hat{\sigma}_{-}-\hat{\sigma}_{+}\hat{a}^{p}\right)\right]\left|\psi(0)\right\rangle\nonumber\\
	&=pg\left\langle \psi(0)\right|\left(\begin{array}{c}
		\left[\omega_{a}\text{\ensuremath{\hat{a}^{\dagger}}}\hat{a}+\omega_{0}\hat{\sigma}_{z}+\frac{U}{2}\hat{a}^{\dagger2}\hat{a}^{2}+\gamma\text{\ensuremath{\hat{a}^{\dagger}}}\hat{a}\hat{\sigma}_{z},\hat{a}^{\dagger p}\hat{\sigma}_{-}-\hat{\sigma}_{+}\hat{a}^{p}\right]\\
		+\left[g(\hat{a}^{\dagger p}\hat{\sigma}_{-}+\hat{\sigma}_{+}\hat{a}^{p}),\hat{a}^{\dagger p}\hat{\sigma}_{-}-\hat{\sigma}_{+}\hat{a}^{p}\right]
	\end{array}\right)\left|\psi(0)\right\rangle\nonumber\\
	&=pg\left\langle \psi(0)\right|\left(\begin{array}{c}
		\left[\omega_{a}\text{\ensuremath{\hat{a}^{\dagger}}}\hat{a}+\omega_{0}\hat{\sigma}_{z}+\frac{U}{2}\hat{a}^{\dagger2}\hat{a}^{2}+\gamma\text{\ensuremath{\hat{a}^{\dagger}}}\hat{a}\hat{\sigma}_{z},\hat{a}^{\dagger p}\hat{\sigma}_{-}-\hat{\sigma}_{+}\hat{a}^{p}\right]\\
		+2g\left(\hat{a}^{p}\hat{a}^{\dagger p}\hat{\sigma}_{+}\hat{\sigma}_{-}-\hat{a}^{\dagger p}\hat{a}^{p}\hat{\sigma}_{-}\hat{\sigma}_{+}\right)
	\end{array}\right)\left|\psi(0)\right\rangle  .\tag{S34}\label{S34}
\end{align}
Thus, for the single-photon JC model, when we prepare the optical field in the vacuum state $\left|0\right\rangle$, we can obtain the probability that the atom is in the excited state at the initial time by measuring the change in the average value of $\hat{a}^{\dagger}\hat{a}$, i.e.,
\begin{equation}
\left\langle \phi_{a}\right|\hat{\sigma}_{+}\hat{\sigma}_{-}\left|\phi_{a}\right\rangle =\left\langle \phi_{r}\right|\hat{\sigma}_{+}\hat{\sigma}_{-}\left|\phi_{r}\right\rangle\lim_{t\rightarrow0}\frac{\Delta\left\langle \text{\ensuremath{\hat{a}^{\dagger}}}\hat{a}\right\rangle \left(t,\left|\phi_{a}\right\rangle \otimes\left|0\right\rangle \right)}{\Delta\left\langle \text{\ensuremath{\hat{a}^{\dagger}}}\hat{a}\right\rangle\text{\ensuremath{\left(t,\left|\phi_{r}\right\rangle \otimes\left|0\right\rangle \right)}} },  \tag{S35}\label{S35}
\end{equation}
where $\left|\phi_{r}\right\rangle$ is a known reference state of the atomic system, e.g., $\left|\phi_{r}\right\rangle=\left|e\right\rangle$.

The results of the $p$-photon Dicke model and the $p$-photon TC model are similar to the $p$-photon Rabi model and the $p$-photon JC model, respectively. We will not continue with the theoretical analysis of the two multi-atom models, but later we will compare the results of these four models two by two using numerical methods. In the following, we investigate whether the results of the $p$-photon Rabi model and the $p$-photon JC model are the same as those known above under the actual dynamical evolution.

\begin{figure}[t]
	\setlength{\abovecaptionskip}{0.cm}
	\setlength{\belowcaptionskip}{-0.cm}
	\centering
	\includegraphics[width=18cm,height=14cm]{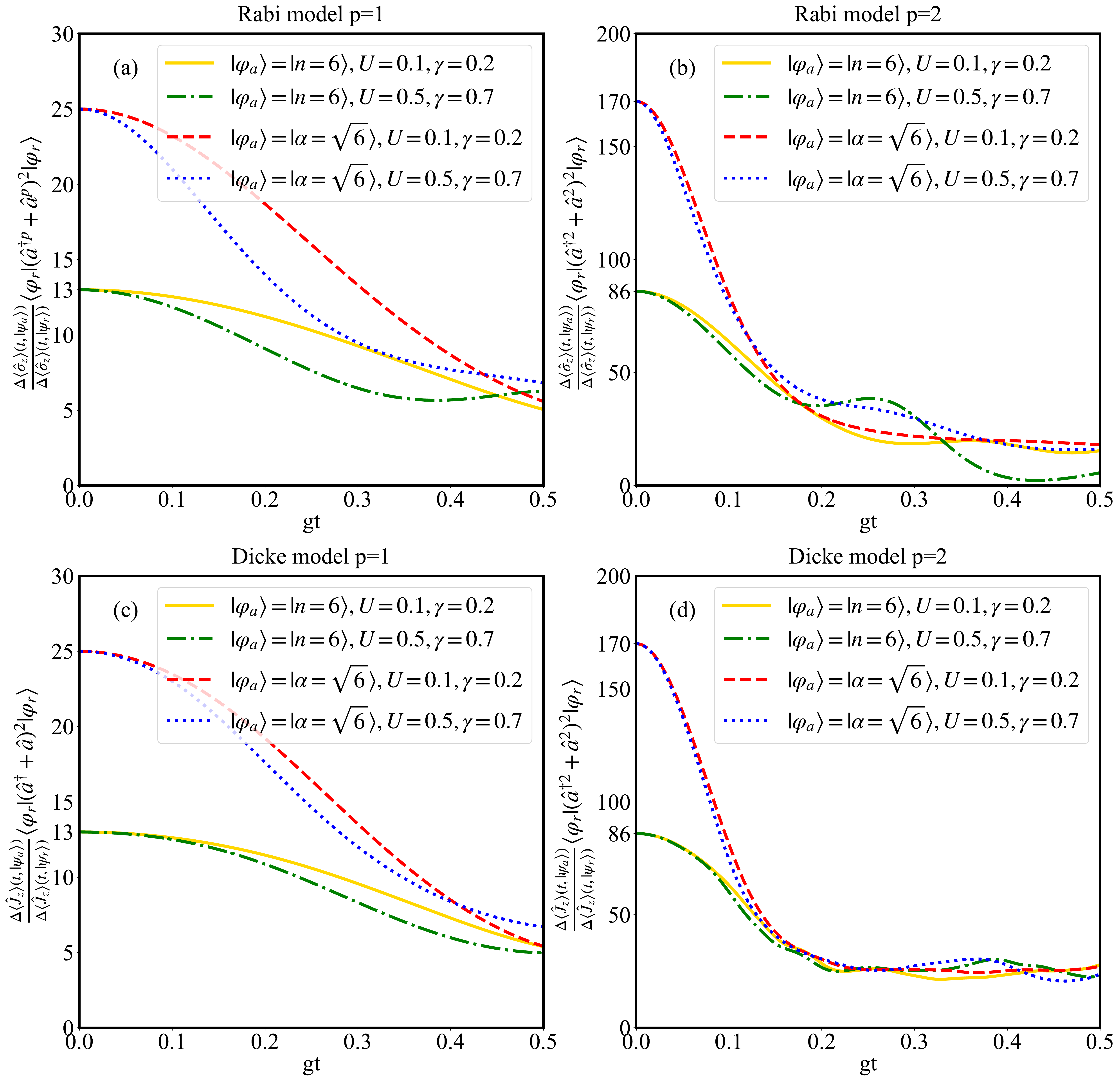}
	\caption{(a) and (b) denote the variation of $\frac{\Delta\langle\hat{\sigma}_{z}\rangle(t,|\psi_{a}\rangle)}{\Delta\langle\hat{\sigma}_{z}\rangle(t,|\psi_{r}\rangle)}\langle\varphi_{r}|(\hat{a}^{\dagger p}+\hat{a}^{p})^2|\varphi_{r}\rangle$ with time $t$ when $p = 1$ and $p = 2$, respectively, in the Rabi model. (c) and (d) represent the change of $\frac{\Delta\langle\hat{J}_{z}\rangle(t,|\psi_{a}\rangle)}{\Delta\langle\hat{J}_{z}\rangle(t,|\psi_{r}\rangle)}\langle\varphi_{r}|(\hat{a}^{\dagger p}+\hat{a}^{p})^2|\varphi_{r}\rangle$ with time $t$ when $p = 1$ and $p = 2$ in the Dicke model, respectively. Here, the initial state in the actual measurement process and the reference state in the Rabi model are $|\psi_{a}\rangle=|\varphi_{a}\rangle\otimes|g\rangle$ and $|\psi_{r}\rangle=|\varphi_{r}\rangle\otimes|g\rangle$, respectively. The initial state in the actual measurement process and the reference state in the Dicke model are $|\psi_{a}\rangle=|\varphi_{a}\rangle\otimes|N/2, -N/2\rangle$ and $|\psi_{r}\rangle=|\varphi_{r}\rangle\otimes|N/2, -N/2\rangle$, respectively, where $|N/2, -N/2\rangle$ is the Dicke state, and the state of the cavity field in the reference state is the number state $|\varphi_{r}\rangle=|n=3\rangle$. In order to numerically calculate the dynamical evolution of the system, we set the unknown cavity field states as the number state $|\varphi_{a}\rangle=|n=6\rangle$ and the coherent state $|\varphi_{a}\rangle=|\alpha=\sqrt{6}\rangle$, respectively. We take the atomic number $N = 10$ in the Dicke model. The values of other parameters are $w_{a}=1$, $w_{0}=w_{a}$, $U=0.1w_{a}$, and $\gamma=0.2w_{a}$.}
	\label{figS2}
\end{figure}

\subsection{Results under actual dynamical evolution}
In this subsection, we study the acquisition of information about the initial cavity field by measuring the change in atomic energy or about the initial atom by measuring the change in the average photon number of the cavity field under the actual dynamical evolution.

First we study the acquisition of information about the optical field by measuring the energy change of the atom in the generalized $p$-photon JC model. The Hamiltonian of $p$-photon JC model is
\begin{equation}
\hat{H}_{JC}=\omega_{a}\text{\ensuremath{\hat{a}^{\dagger}}}\hat{a}+\frac{\omega_{0}}{2}\hat{\sigma}_{z}+g(\hat{a}^{\dagger p}\hat{\sigma}_{-}+\hat{\sigma}_{+}\hat{a}^{p})+\frac{U}{2}\hat{a}^{\dagger2}\hat{a}^{2}+\gamma\text{\ensuremath{\hat{a}^{\dagger}}}\hat{a}\hat{\sigma}_{z}.  \tag{S36}\label{S36}
\end{equation}
We first take the state of the unknown cavity field as the object of study and obtain information about the cavity field by measuring the change of the atomic energy. Here, we assume that the cavity field is in an unknown state $\left|\varphi_{a}(0)\right\rangle$ and the atom is in the ground state $\left|g\right\rangle$, that is, the initial state of the system is $\left|\psi(0)\right\rangle=\left|\varphi_{a}(0)\right\rangle\otimes\left|g\right\rangle=\sum_{n=0}^{\infty}c_{n}|g,n\rangle$, where $c_{n}=\left\langle n|\varphi_{a}(0)\right\rangle$ is the inner product of the number state  $\left|n\right\rangle$ and the cavity field state $\left|\varphi_{a}(0)\right\rangle$. Since the total excitation number is conserved (the total excitation number operator $\hat{N}_{e}=\hat{a}^{\dagger}\hat{a}+p\left|e\right\rangle\left\langle e\right|$ satisfies the condition $[\hat{N}_{e}, \hat{H}_{JC}]=0$), we can assume that the state of the system at time $t$ has the following form
\begin{equation}
\left|\psi(t)\right\rangle=\sum_{n=0}^{\infty}c_{n}[C_{n-p}^{e}(t)|e,n-p\rangle+C_{n}^{g}(t)|g,n\rangle],   \tag{S37}\label{S37}
\end{equation}
where $C_{n-p}^{e}(0)=0$ and $C_{n}^{g}(0)=1$. Substituting Eq.~(\ref{S36}) and Eq.~(\ref{S37}) into Schr\"{o}dinger Equation, we can obatin two differential equations
\begin{align}
	i\frac{d}{dt}C_{n-p}^{e}(t)&=AC_{n-p}^{e}(t)+BC_{n}^{g}(t),  \tag{S38}\label{S38}\\
	i\frac{d}{dt}C_{n}^{g}(t)&=BC_{n-p}^{e}(t)+DC_{n}^{g}(t),   \tag{S39}\label{S39}
\end{align} 
where 
\begin{align}
	A&=\frac{\omega_{0}}{2}+\omega_{a}(n-p)+\frac{U}{2}(n-p)(n-p-1)+\gamma(n-p),   \tag{S40}\label{S40}\\
	B&=g\sqrt{n!/(n-p)!} ,\tag{S41}\label{S41}\\
	D&=-\frac{\omega_{0}}{2}+\omega_{a}n-\frac{U}{2}n(n-1)-\gamma n  .\tag{S42}\label{S42}
\end{align}
The solutions of the two differential equations above are
\begin{align}
	C_{n-p}^{e}(t)&=z_{1}\left[\cos(x_{2}t)-\cos(x_{1}t)\right]-i\left(z_{1}\left[\sin(x_{2}t)-\sin(x_{1}t)\right]\right) ,\tag{S43}\label{S43}\\
	C_{n}^{g}(t)&=y_{1}\cos(x_{1}t)-y_{2}\cos(x_{2}t)-i\left[y_{1}\sin(x_{1}t)-y_{2}\sin(x_{2}t)\right] , \tag{S44}\label{S44}
\end{align}
where
\begin{align}
	x_{1}&=\frac{(A+D)-\sqrt{(A-D)^{2}+4B^{2}}}{2}, x_{2}=\frac{(A+D)+\sqrt{(A-D)^{2}+4B^{2}}}{2}, \tag{S45}\label{S45}\\
	y_{1}&=\frac{A-x_{1}}{x_{2}-x_{1}}, y_{2}=\frac{A-x_{2}}{x_{2}-x_{1}} , \tag{S46}\label{S46} \\
	z_{1}&=\frac{B}{x_{2}-x_{1}} . \tag{S47}\label{S47}
\end{align}
\begin{figure}[t!]
	\setlength{\abovecaptionskip}{0.cm}
	\setlength{\belowcaptionskip}{-0.cm}
	\centering
	\includegraphics[width=18cm,height=14cm]{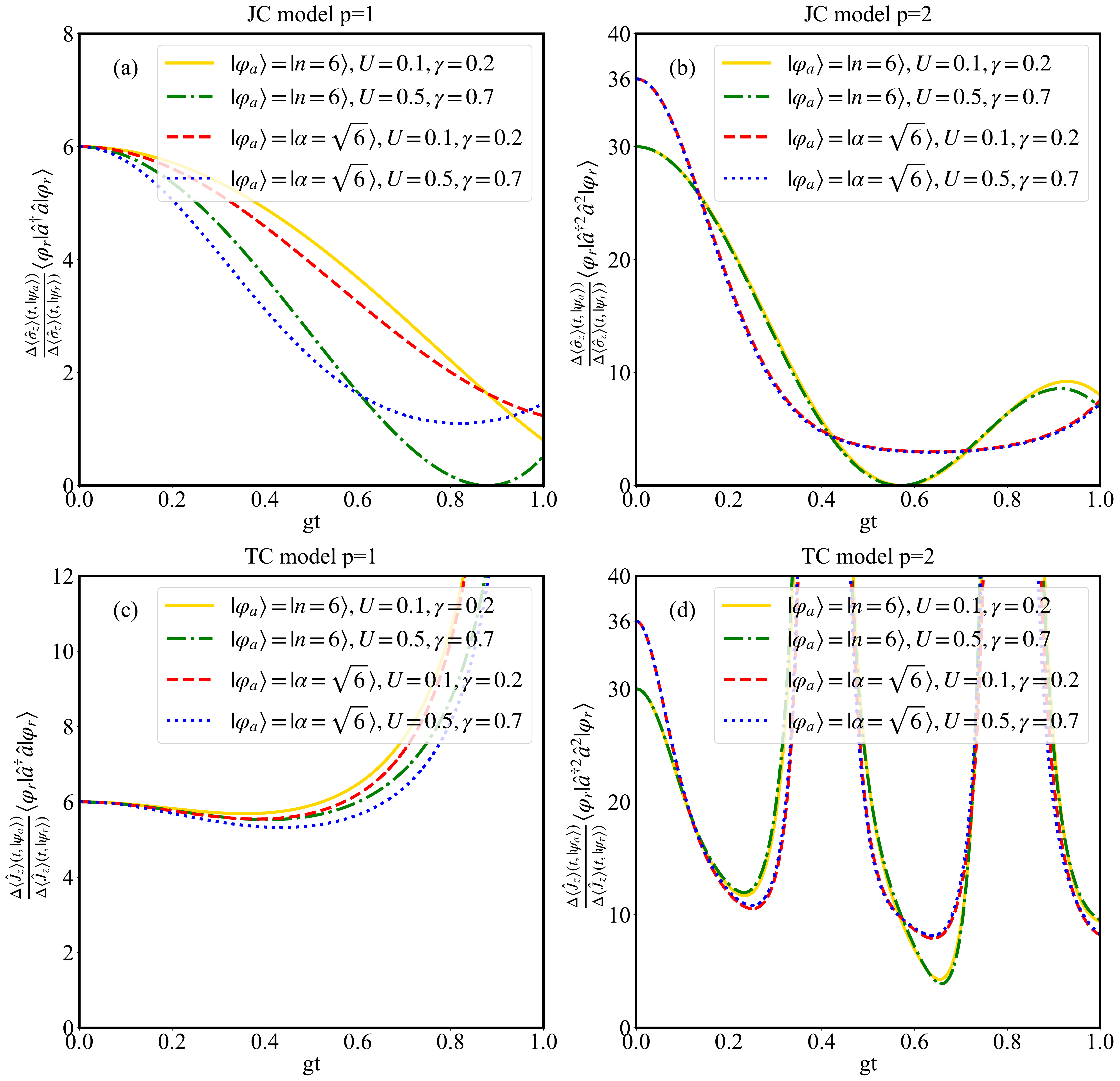}
	\caption{(a) and (b) denote the variation of $\frac{\Delta\langle\hat{\sigma}_{z}\rangle(t,|\psi_{a}\rangle)}{\Delta\langle\hat{\sigma}_{z}\rangle(t,|\psi_{r}\rangle)}\langle\varphi_{r}|\hat{a}^{\dagger p}\hat{a}^{p}|\varphi_{r}\rangle$ with time $t$ when $p = 1$ and $p = 2$, respectively, in the JC model. (c) and (d) represent the change of $\frac{\Delta\langle\hat{J}_{z}\rangle(t,|\psi_{a}\rangle)}{\Delta\langle\hat{J}_{z}\rangle(t,|\psi_{r}\rangle)}\langle\varphi_{r}|\hat{a}^{\dagger p}\hat{a}^{p}|\varphi_{r}\rangle$ with time $t$ when $p = 1$ and $p = 2$ in the TC model, respectively. The other parameters are the same as in Fig.~\ref{figS2}.}
	\label{figS3}
\end{figure}
When the system evolves with another known reference initial state $\left|\psi_{r}(0)\right\rangle=\left|\varphi_{r}(0)\right\rangle\otimes\left|g\right\rangle=\sum_{n=0}^{\infty}d_{n}|g,n\rangle$, where $d_{n}=\left\langle n|\varphi_{r}(0)\right\rangle$, the state of the system at time $t$ is
\begin{equation}
\left|\psi_{r}(t)\right\rangle=\sum_{n=0}^{\infty}d_{n}[C_{n-p}^{e}(t)|e,n-p\rangle+C_{n}^{g}(t)|g,n\rangle] .  \tag{S48}\label{S48}
\end{equation}
In these two different initial states, the ratio of the energy gained by the atom from the cavity field is
\begin{equation}
\frac{\Delta\langle\hat{\sigma}_{z}\rangle(|\psi(0)\rangle)}{\Delta\langle\hat{\sigma}_{z}\rangle(|\phi_{r}(0)\rangle)}=\frac{\sum_{n=0}^{\infty}\left|c_{n}\right|^{2}\left(\left|C_{n-p}^{e}(t)\right|^{2}-\left|C_{n}^{g}(t)\right|^{2}\right)+1}{\sum_{m=0}^{\infty}\left|d_{m}\right|^{2}\left(\left|C_{m-p}^{e}(t)\right|^{2}-\left|C_{m}^{g}(t)\right|^{2}\right)+1}. \tag{S49}\label{S49}
\end{equation}
Since $\lim_{t\rightarrow0}[\sum_{n=0}^{\infty}\left|c_{n}\right|^{2}(\left|C_{n-p}^{e}(t)\right|^{2}-\left|C_{n}^{g}(t)\right|^{2})+1]=0$, $\lim_{t\rightarrow0}\frac{d}{dt}\left|C_{n-p}^{e}(t)\right|^{2}=\lim_{t\rightarrow0}\frac{d}{dt}\left|C_{n}^{g}(t)\right|^{2}=0$, and $\lim_{t\rightarrow0}\frac{d^{2}}{dt^{2}}\left|C_{n-p}^{e}(t)\right|^{2}=-\lim_{t\rightarrow0}\frac{d^{2}}{dt^{2}}\left|C_{n}^{g}(t)\right|^{2}=2B^{2}$, we can obtain
\begin{equation}
\lim_{t\rightarrow0}\frac{\Delta\langle\hat{\sigma}_{z}\rangle(|\psi(0)\rangle)}{\Delta\langle\hat{\sigma}_{z}\rangle(|\psi_{r}(0)\rangle)}=\frac{\sum_{n=0}^{\infty}\left|c_{n}\right|^{2}[n!/(n-p)!]}{\sum_{m=0}^{\infty}\left|d_{m}\right|^{2}[m!/(m-p)!]}=\frac{\langle\varphi_{a}(0)|\hat{a}^{\dagger p}\hat{a}^{p}|\varphi_{a}(0)\rangle}{\langle\varphi_{r}(0)|\hat{a}^{\dagger p}\hat{a}^{p}|\varphi_{r}(0)\rangle},  \tag{S50}\label{S50}
\end{equation}
where $\langle\varphi_{a}(0)|\hat{a}^{\dagger p}\hat{a}^{p}|\varphi_{a}(0)\rangle$ and $\langle\varphi_{r}(0)|\hat{a}^{\dagger p}\hat{a}^{p}|\varphi_{r}(0)\rangle$ denote the $p$-order correlation functions of the arbitrary state $|\varphi_{a}(0)\rangle$ of the cavity field and the reference state $|\varphi_{r}(0)\rangle$ of the cavity field, respectively. This result is consistent with Eq.~(\ref{S25}). Obviously, we can obtain information about the unknown cavity field by measuring the energy change of the atom with the aid of the known reference cavity field, i.e.,
\begin{equation}
\langle\varphi_{a}(0)|\hat{a}^{\dagger p}\hat{a}^{p}|\varphi_{a}(0)\rangle=\langle\varphi_{r}(0)|\hat{a}^{\dagger p}\hat{a}^{p}|\varphi_{r}(0)\rangle\lim_{t\rightarrow0}\frac{\Delta\langle\hat{\sigma}_{z}\rangle(t,|\psi(0)\rangle)}{\Delta\langle\hat{\sigma}_{z}\rangle(t,|\psi_{r}(0)\rangle)} . \tag{S51}\label{S51}
\end{equation}
In Fig.~\ref{figS1}, we give a schematic of obtaining information about the cavity field by measuring the variation of the atomic energy. At the initial moment, we prepare the atom in the ground state, i.e., $\left|\phi(0)\right\rangle=\left|g\right\rangle$. Then, the atom enter the cavity field and leave the cavity field after a instantaneous interaction with the cavity field, at which point we only need to measure the energy of the atom leaving the cavity field to obtain information about the cavity field. The advantages of this indirect measurement are mainly as follows: firstly, the short time interaction of the atom with the cavity field does not change the cavity field state much and this measurement does not cause wave function collapse, secondly, the  measurement does not require any additional manipulation of the cavity field other than the dipole interaction, and finally, the possible additional photon-photon interaction and photon-atom interaction during the interaction has no influence on the measurement results.

In Fig.~\ref{figS2}, for the extended Rabi and Dicke models, we plot the variation of $\frac{\Delta\langle\hat{\sigma}_{z}\rangle(t,|\psi_{a}\rangle)}{\Delta\langle\hat{\sigma}_{z}\rangle(t,|\psi_{r})\rangle}\langle\varphi_{r}|(\hat{a}^{\dagger p}+\hat{a}^{p})^2|\varphi_{r}\rangle$ and $\frac{\Delta\langle\hat{J}_{z}\rangle(t,|\psi_{a}\rangle)}{\Delta\langle\hat{J}_{z}\rangle(t,|\psi_{r}\rangle)}\langle\varphi_{r}|(\hat{a}^{\dagger p}+\hat{a}^{p})^2|\varphi_{r}\rangle$ with time $t$ in the evolution of the actual dynamics when $p$ takes different values, respectively. The Hamiltonian of the extended $p$-photon Dicke model is as follows
\begin{equation}
\hat{H}_{D}=\omega_{a}\text{\ensuremath{\hat{a}^{\dagger}}}\hat{a}+\omega_{0}\hat{J}_{z}+g(\hat{a}^{\dagger p}+\hat{a}^{p})(\hat{J}_{-}+\hat{J}_{+})+\frac{U}{2}\hat{a}^{\dagger2}\hat{a}^{2}+\gamma\text{\ensuremath{\hat{a}^{\dagger}}}\hat{a}
\hat{J}_{z}   , \tag{S52}\label{S52}
\end{equation}
where $\hat{J}_{\alpha}$ ($\alpha=x,y,z$) is the collective angular momentum operator for the spin ensemble consisting of $N$ identical two-level atoms; these operators $\hat{J}_{x}, \hat{J}_{y}, \hat{J}_{z}$ satisfy the commutation relation of SU(2) algebra and $\hat{J}_{\pm}=\hat{J}_{x}\pm i\hat{J}_{y}$. From Fig.~\ref{figS2}, we can see that for different initial states of the cavity field, when $t\rightarrow0$, the following relationship holds in the Rabi and Dicke models 
\begin{align}
	\left\langle\varphi_{a}(0)\right|(\hat{a}^{\dagger p}+\hat{a}^{p})^{2}\left|\varphi_{a}(0)\right\rangle&=\lim_{t\rightarrow0}\frac{\Delta\left\langle \hat{\sigma}_{z}\right\rangle(t,\left|\psi(0)\right\rangle ) }{\Delta\left\langle \hat{\sigma}_{z}\right\rangle(t,\left|\psi_{r}(0)\right\rangle ) }\left\langle\varphi_{r}(0)\right|(\hat{a}^{\dagger p}+\hat{a}^{p})^{2}\left|\varphi_{r}(0)\right\rangle \nonumber \\
	&=\lim_{t\rightarrow0}\frac{\Delta\langle \hat{J}_{z}\rangle(t,\left|\psi(0)\right\rangle ) }{\Delta\langle \hat{J}_{z}\rangle(t,\left|\psi_{r}(0)\right\rangle ) }\left\langle\varphi_{r}(0)\right|(\hat{a}^{\dagger p}+\hat{a}^{p})^{2}\left|\varphi_{r}(0)\right\rangle ,  p=1,2.\tag{S53}\label{S53}  
\end{align}

Similarly, we can see from Fig.~\ref{figS3} that for different initial states of the cavity field, when $t\rightarrow0$, the following relationship holds in the extended JC model and the extended TC model.
\begin{align}
	\left\langle\varphi_{a}(0)\right|\hat{a}^{\dagger p}\hat{a}^{p}\left|\varphi_{a}(0)\right\rangle&=\lim_{t\rightarrow0}\frac{\Delta\left\langle \hat{\sigma}_{z}\right\rangle(t,\left|\psi(0)\right\rangle ) }{\Delta\left\langle \hat{\sigma}_{z}\right\rangle(t,\left|\psi_{r}(0)\right\rangle ) }\left\langle\varphi_{r}(0)\right|\hat{a}^{\dagger p}\hat{a}^{p}\left|\varphi_{r}(0)\right\rangle \nonumber \\
	&=\lim_{t\rightarrow0}\frac{\Delta\langle \hat{J}_{z}\rangle(t,\left|\psi(0)\right\rangle ) }{\Delta\langle \hat{J}_{z}\rangle(t,\left|\psi_{r}(0)\right\rangle ) }\left\langle\varphi_{r}(0)\right|\hat{a}^{\dagger p}\hat{a}^{p}\left|\varphi_{r}(0)\right\rangle ,  p=1,2.\tag{S54}\label{S54}  
\end{align}
where the Hamiltonian of the extended $p$-photon TC model is
\begin{equation}
\hat{H}_{TC}=\omega_{a}\text{\ensuremath{\hat{a}^{\dagger}}}\hat{a}+\omega_{0}\hat{J}_{z}+g(\hat{a}^{\dagger p}\hat{J}_{-}+\hat{J}_{+}\hat{a}^{p})+\frac{U}{2}\hat{a}^{\dagger2}\hat{a}^{2}+\gamma\text{\ensuremath{\hat{a}^{\dagger}}}\hat{a}\hat{J}_{z}. \tag{S55}\label{S55}
\end{equation}
This interesting phenomenon in Fig.~\ref{figS3} has also appeared before in Ref. \cite{PhysRevA.104.043706}.

In this section, we obtain results consistent with the previous theoretical analysis by considering the actual dynamical evolution of the different models. All the data obtained from the numerical calculations in Fig.~\ref{figS2} and Fig.~\ref{figS3} were performed by QuTIP \cite{JOHANSSON20121760}, a quantum toolbox.

\section{The instantaneous indirect measurement of quantum fisher information}
Quantum Fisher information is a core quantity in quantum metrology. Therefore, how to find the maximum quantum Fisher information of the system and how to implement the measurement of the quantum Fisher information are important research elements \cite{PhysRevA.88.060101,RevModPhys.90.035005,PhysRevA.105.023718,liu2019quantum,PhysRevLett.126.120503}. In this section, similar to the first section, we look for instantaneous measurements of the quantum Fisher information for a quantum system containing some parameter $\lambda$. 

We consider a Hamiltonian $\hat{H}(\lambda)$ with parameter $\lambda$ and its initial state $\left|\Psi(0)\right\rangle $, then the state of the system at time $t$ is
\begin{align}
	\left|\Psi(t)\right\rangle &=\exp(-i\hat{H}(\lambda)t)\left|\Psi(0)\right\rangle \nonumber\\
	&=\sum_{n=0}^{\infty}\frac{(-it)^{n}}{n!}\left[\hat{H}(\lambda)\right]^{n}\left|\Psi(0)\right\rangle    . \tag{S56}\label{S56}
\end{align}
The quantum Fisher information of the parameter $\lambda$ in the pure state $\left|\Psi(t)\right\rangle$  is
\begin{align}
	F_{\lambda}(t,\left|\Psi(0)\right\rangle )&=4\left[\langle\partial_{\lambda}\Psi(t)\left|\partial_{\lambda}\Psi(t)\right\rangle -\left|\langle\Psi(t)\left|\partial_{\lambda}\Psi(t)\right\rangle \right|^{2}\right]\nonumber\\
	&=4\left[\langle\partial_{\lambda}\Psi(t)\left|\partial_{\lambda}\Psi(t)\right\rangle -\left[\frac{\left(\langle\Psi(t)\left|\partial_{\lambda}\Psi(t)\right\rangle +\langle\partial_{\lambda}\Psi(t)\left|\Psi(t)\right\rangle \right)^{2}}{4}-\frac{\left(\langle\Psi(t)\left|\partial_{\lambda}\Psi(t)\right\rangle -\langle\partial_{\lambda}\Psi(t)\left|\Psi(t)\right\rangle \right)^{2}}{4}\right]\right], \tag{S57}\label{S57}
\end{align}
where
\begin{equation}
\left|\partial_{\lambda}\Psi(t)\right\rangle =\sum_{n=0}^{\infty}\frac{(-it)^{n+1}}{(n+1)!}\frac{\partial\left[\hat{H}(\lambda)\right]^{n+1}}{\partial\lambda}\left|\Psi(0)\right\rangle.   \tag{S58}\label{S58}
\end{equation}
When the system evolves in another initial state $\left|\Psi_{r}(0)\right\rangle$, then the state at time $t$ is
\begin{align}
	\left|\Psi_{r}(t)\right\rangle &=\exp(-i\hat{H}(\lambda)t)\left|\Psi_{r}(0)\right\rangle \nonumber\\
	&=\sum_{n=0}^{\infty}\frac{(-it)^{n}}{n!}\left[\hat{H}(\lambda)\right]^{n}\left|\Psi_{r}(0)\right\rangle  , \tag{S59}\label{S59}
\end{align}
and the quantum Fisher information of the parameter $\lambda$ for the state $\left|\Psi_{r}(t)\right\rangle$ (we call it a reference state) is
\begin{equation}
F_{\lambda}(t,\left|\Psi_{r}(0)\right\rangle )=4\left[\langle\partial_{\lambda}\Psi_{r}(t)\left|\partial_{\lambda}\Psi_{r}(t)\right\rangle -\left|\langle\Psi_{r}(t)\left|\partial_{\lambda}\Psi_{r}(t)\right\rangle \right|^{2}\right],\tag{S60}\label{S60}
\end{equation}
where
\begin{equation}
\left|\partial_{\lambda}\Psi_{r}(t)\right\rangle =\sum_{n=0}^{\infty}\frac{(-it)^{n+1}}{(n+1)!}\frac{\partial\left[\hat{H}(\lambda)\right]^{n+1}}{\partial\lambda}\left|\Psi_{r}(0)\right\rangle .  \tag{S61}\label{S61}
\end{equation}
When $t\rightarrow0$, the ratio of the quantum Fisher information of the parameter for different initial states is
\begin{equation}
\lim_{t\rightarrow0}\frac{F_{\lambda}(t,\left|\Psi(0)\right\rangle )}{F_{\lambda}(t,\left|\Psi_{r}(0)\right\rangle )}=\frac{\langle\partial_{\lambda}\Psi(t)\left|\partial_{\lambda}\Psi(t)\right\rangle -\left|\langle\Psi(t)\left|\partial_{\lambda}\Psi(t)\right\rangle \right|^{2}}{\langle\partial_{\lambda}\Psi_{r}(t)\left|\partial_{\lambda}\Psi_{r}(t)\right\rangle -\left|\langle\Psi_{r}(t)\left|\partial_{\lambda}\Psi_{r}(t)\right\rangle \right|^{2}}. \tag{S62}\label{S62}
\end{equation}
Since
\begin{align}
	\lim_{t\rightarrow0}\left|\partial_{\lambda}\Psi(t)\right\rangle &=\sum_{n=0}^{\infty}\frac{(-it)^{n+1}}{(n+1)!}\frac{\partial\left[\hat{H}(\lambda)\right]^{n+1}}{\partial\lambda}\left|\Psi(0)\right\rangle =0, \tag{S63}\label{S63}\\
	\lim_{t\rightarrow0}\left|\partial_{\lambda}\Psi_{r}(t)\right\rangle &=\sum_{n=0}^{\infty}\frac{(-it)^{n+1}}{(n+1)!}\frac{\partial\left[\hat{H}(\lambda)\right]^{n+1}}{\partial\lambda}\left|\Psi_{r}(0)\right\rangle =0. \tag{S64}\label{S64}
\end{align}
Then
\begin{align}
	\lim_{t\rightarrow0}F_{\lambda}(t,\left|\Psi(0)\right\rangle )&=4\lim_{t\rightarrow0}\left[\langle\partial_{\lambda}\Psi(t)\left|\partial_{\lambda}\Psi(t)\right\rangle -\left|\langle\Psi(t)\left|\partial_{\lambda}\Psi(t)\right\rangle \right|^{2}\right]=0 ,\tag{S65}\label{S65}\\
	\lim_{t\rightarrow0}F_{\lambda}(t,\left|\Psi_{r}(0)\right\rangle )&=4\lim_{t\rightarrow0}\left[\langle\partial_{\lambda}\Psi_{r}(t)\left|\partial_{\lambda}\Psi_{r}(t)\right\rangle -\left|\langle\Psi_{r}(t)\left|\partial_{\lambda}\Psi_{r}(t)\right\rangle \right|^{2}\right]=0 ,\tag{S66}\label{S66} \\
	\lim_{t\rightarrow0}\frac{dF_{\lambda}(t,\left|\Psi(0)\right\rangle )}{dt}&=0, \tag{S67}\label{S67}\\
	\lim_{t\rightarrow0}\frac{dF_{\lambda}(t,\left|\Psi_{r}(0)\right\rangle )}{dt}&=0.\tag{S68}\label{S68}
\end{align}
Then, we should consider the second-order derivative of $F_{\lambda}(t,\left|\Psi(0)\right\rangle )$ with respect to $t$, i.e.,
\begin{align}
	\lim_{t\rightarrow0}\frac{d^{2}F_{\lambda}(t,\left|\Psi(0)\right\rangle )}{dt^{2}}&=4\lim_{t\rightarrow0}\Bigg[2\frac{d\left(\langle\partial_{\lambda}\Psi(t)|\right)}{dt}\frac{d\left|\partial_{\lambda}\Psi(t)\right\rangle }{dt}-\frac{1}{2}\left(\langle\Psi(t)|\frac{d\left|\partial_{\lambda}\Psi(t)\right\rangle }{dt}+\frac{d\langle\partial_{\lambda}\Psi(t)|}{dt}\left|\Psi(t)\right\rangle \right)^{2}\nonumber\\
	&+\frac{1}{2}\left(\langle\Psi(t)|\frac{d\left|\partial_{\lambda}\Psi(t)\right\rangle }{dt}-\frac{d\langle\partial_{\lambda}\Psi(t)|}{dt}\left|\Psi(t)\right\rangle \right)^{2}\Bigg]. \tag{S69}\label{S69}
\end{align}
Since
\begin{align}
	\lim_{t\rightarrow0}\frac{d\left|\partial_{\lambda}\Psi(t)\right\rangle }{dt}&=\lim_{t\rightarrow0}\sum_{n=0}^{\infty}\frac{-i(-it)^{n}}{n!}\frac{\partial\left[\hat{H}(\lambda)\right]^{n+1}}{\partial\lambda}\left|\Psi(0)\right\rangle \nonumber\\
	&=-i\frac{\partial\hat{H}(\lambda)}{\partial\lambda}\left|\Psi(0)\right\rangle ,\tag{S70}\label{S70}
\end{align}
then
\begin{align}
	\lim_{t\rightarrow0}\frac{d^{2}F_{\lambda}(\left|\Psi(0)\right\rangle )}{dt^{2}}&=8\Bigg[\left\langle \Psi(0)\right|\left[\frac{\partial\hat{H}(\lambda)}{\partial\lambda}\right]^{\dagger}\frac{\partial\hat{H}(\lambda)}{\partial\lambda}\left|\Psi(0)\right\rangle +\left(\textbf{Im}(\left\langle \Psi(0)\right|\frac{\partial\hat{H}(\lambda)}{\partial\lambda}\left|\Psi(0)\right\rangle )\right)^{2} \nonumber\\
	&-\left(\textbf{Re}(\left\langle \Psi(0)\right|\frac{\partial\hat{H}(\lambda)}{\partial\lambda}\left|\Psi(0)\right\rangle )\right)^{2}\Bigg].\tag{S71}\label{S71}
\end{align}
Since $\hat{H}(\lambda)$ is a Hermitian operator, $\partial\hat{H}(\lambda)/\partial\lambda$ is also a Hermitian operator for the real parameter $\lambda$. We give the proof of this conclusion below. 

Since $\hat{H}(\lambda)$ is a Hermitian operator, then $\langle\varphi_{1}|\hat{H}(\lambda)|\varphi_{2}\rangle=\langle\hat{H}(\lambda)\varphi_{1}|\varphi_{2}\rangle$, where $\langle\hat{H}(\lambda)\varphi_{1}|$ denote $(\hat{H}(\lambda)|\varphi_{1}\rangle)^{\dagger}$,  $|\varphi_{1}\rangle$ and $|\varphi_{2}\rangle$ are two arbitrary states, and since 
\begin{align}
	\left\langle \varphi_{1}\right|\frac{\partial\hat{H}(\lambda)}{\partial\lambda}\left|\varphi_{2}\right\rangle &=\left\langle \varphi_{1}\right|\frac{\partial}{\partial\lambda}\left(\hat{H}(\lambda)\left|\varphi_{2}\right\rangle \right)\nonumber\\
	&=\frac{\partial}{\partial\lambda}\left(\left\langle \varphi_{1}\right|\hat{H}(\lambda)\left|\varphi_{2}\right\rangle \right)-\left\langle \frac{\partial}{\partial\lambda^{*}}\varphi_{1}\right|\hat{H}(\lambda)\left|\varphi_{2}\right\rangle \nonumber\\
	&=\frac{\partial}{\partial\lambda}\left(\left\langle \varphi_{1}\right|\hat{H}(\lambda)\left|\varphi_{2}\right\rangle \right)\nonumber\\
	&=\left\langle \frac{\partial\hat{H}(\lambda)}{\partial\lambda^{*}}\varphi_{1}\right|\varphi_{2}\rangle,	\tag{S72}\label{S72}
\end{align}
if $\lambda$ is a real parameter, then $\left\langle \varphi_{1}\right|\frac{\partial\hat{H}(\lambda)}{\partial\lambda}\left|\varphi_{2}\right\rangle =\left\langle \frac{\partial\hat{H}(\lambda)}{\partial\lambda}\varphi_{1}\right|\varphi_{2}\rangle$, that is,  $\frac{\partial\hat{H}(\lambda)}{\partial\lambda}$ is a Hermitian operator. 

Finally, we can obtain the following expression
\begin{equation}
\lim_{t\rightarrow0}\frac{F_{\lambda}(t,\left|\Psi(0)\right\rangle )}{F_{\lambda}(t,\left|\Psi_{r}(0)\right\rangle )}=\frac{\left\langle \Psi(0)\right|\left[\frac{\partial\hat{H}(\lambda)}{\partial\lambda}\right]^{2}\left|\Psi(0)\right\rangle -\left(\left\langle \Psi(0)\right|\frac{\partial\hat{H}(\lambda)}{\partial\lambda}\left|\Psi(0)\right\rangle \right)^{2}}{\left\langle \Psi_{r}(0)\right|\left[\frac{\partial\hat{H}(\lambda)}{\partial\lambda}\right]^{2}\left|\Psi_{r}(0)\right\rangle -\left(\left\langle \Psi_{r}(0)\right|\frac{\partial\hat{H}(\lambda)}{\partial\lambda}\left|\Psi_{r}(0)\right\rangle \right)^{2}}.\tag{S73}\label{S73}
\end{equation}
From the above equation, when $t \rightarrow 0$, we can see that the ratio of the quantum Fisher information with respect to a parameter in different initial states is actually the ratio of the variance of the first-order derivativeof the Hamiltonian with respect to the parameter in different initial states. In the following, we use the $p$-photon Rabi model and the $p$-photon JC model as examples to find the instantaneous quantum Fisher information for the coupling strength $g$ in different initial states.

\subsection{Applications of the instantaneous indirect measurement of quantum fisher information}
For an extended $p$-photon Rabi model and an extended $p$-photon JC model, their Hamiltonians are as follows
\begin{align}
	\hat{H}_{R}&=\omega_{a}\text{\ensuremath{\hat{a}^{\dagger}}}\hat{a}+\frac{\omega_{0}}{2}\hat{\sigma}_{z}+g(\hat{a}^{\dagger p}+\hat{a}^{p})(\hat{\sigma}_{-}+\hat{\sigma}_{+})+\frac{U}{2}\hat{a}^{\dagger2}\hat{a}^{2}+\gamma\text{\ensuremath{\hat{a}^{\dagger}}}\hat{a}\hat{\sigma}_{z} ,\tag{S74}\label{S74} \\
	\hat{H}_{JC}&=\omega_{a}\text{\ensuremath{\hat{a}^{\dagger}}}\hat{a}+\frac{\omega_{0}}{2}\hat{\sigma}_{z}+g(\hat{a}^{\dagger p}\hat{\sigma}_{-}+\hat{\sigma}_{+}\hat{a}^{p})+\frac{U}{2}\hat{a}^{\dagger2}\hat{a}^{2}+\gamma\text{\ensuremath{\hat{a}^{\dagger}}}\hat{a}\hat{\sigma}_{z} , \tag{S75}\label{S75}
\end{align}
their first-order derivatives with respect to the parameter $g$ are
\begin{align}
	\frac{\partial\hat{H}_{R}}{\partial g}&=(\hat{a}^{\dagger p}+\hat{a}^{p})(\hat{\sigma}_{-}+\hat{\sigma}_{+}), \tag{S76}\label{S76}\\
	\frac{\partial\hat{H}_{JC}}{\partial g}&=\hat{a}^{\dagger p}\hat{\sigma}_{-}+\hat{\sigma}_{+}\hat{a}^{p}.\tag{S77}\label{S77}
\end{align}

For the Rabi model, when we prepare the initial state of the system in $|\psi(0)\rangle=|\varphi_{a}(0)\rangle\otimes|g\rangle$ (or $\otimes|e\rangle$) and the reference state is $|\psi_{r}(0)\rangle=|\varphi_{r}(0)\rangle\otimes|g\rangle$ (or $\otimes|e\rangle$), i.e., the atom in the ground state $|g\rangle$ or excited state $|e\rangle$, the initial states of the target and reference cavity fields to be an arbitrary unknown state $|\varphi_{a}(0)\rangle$ and a known state $|\varphi_{r}(0)\rangle$, respectively. From Eq.~(\ref{S73}), we get
\begin{equation}
\lim_{t\rightarrow0}\frac{F_{\lambda}(t,|\psi(0)\rangle )}{F_{\lambda}(t,|\psi_{r}(0)\rangle)}=\frac{\left\langle\varphi_{a}(0)\right|(\hat{a}^{\dagger p}+\hat{a}^{p})^{2}\left|\varphi_{a}(0)\right\rangle}{\left\langle\varphi_{r}(0)\right|(\hat{a}^{\dagger p}+\hat{a}^{p})^{2}\left|\varphi_{r}(0)\right\rangle}=\lim_{t\rightarrow0}\frac{\Delta\langle\hat{\sigma}_{z}\rangle(|\psi(0)\rangle)}{\Delta\langle\hat{\sigma}_{z}\rangle(|\psi_{r}(0)\rangle)}.\tag{S78}\label{S78}
\end{equation}

For the JC model, when we prepare the initial state of the system in $|\psi(0)\rangle=|\varphi_{a}(0)\rangle\otimes|g\rangle$ and the reference state is $|\phi_{r}(0)\rangle=|\varphi_{r}(0)\rangle\otimes|g\rangle$, then
\begin{equation}
\lim_{t\rightarrow0}\frac{F_{\lambda}(t,|\psi(0)\rangle )}{F_{\lambda}(t,|\psi_{r}(0)\rangle)}=\frac{\left\langle\varphi_{a}(0)\right|\hat{a}^{\dagger p}\hat{a}^{p}\left|\varphi_{a}(0)\right\rangle}{\left\langle\varphi_{r}(0)\right|\hat{a}^{\dagger p}\hat{a}^{p}\left|\varphi_{r}(0)\right\rangle}=\lim_{t\rightarrow0}\frac{\Delta\langle\hat{\sigma}_{z}\rangle(|\psi(0)\rangle)}{\Delta\langle\hat{\sigma}_{z}\rangle(|\psi_{r}(0)\rangle)},\tag{S79}\label{S79}
\end{equation}
when we prepare the initial state of the system in $|\psi(0)\rangle=|\varphi_{a}(0)\rangle\otimes|e\rangle$ and the reference state is $|\psi_{r}(0)\rangle=|\varphi_{r}(0)\rangle\otimes|e\rangle$, then
\begin{equation}
\lim_{t\rightarrow0}\frac{F_{\lambda}(t,|\psi(0)\rangle )}{F_{\lambda}(t,|\psi_{r}(0)\rangle)}=\frac{\left\langle\varphi_{a}(0)\right|\hat{a}^{p}\hat{a}^{\dagger p}\left|\varphi_{a}(0)\right\rangle}{\left\langle\varphi_{r}(0)\right|\hat{a}^{p}\hat{a}^{\dagger p}\left|\varphi_{r}(0)\right\rangle}=\lim_{t\rightarrow0}\frac{\Delta\langle\hat{\sigma}_{z}\rangle(|\psi(0)\rangle)}{\Delta\langle\hat{\sigma}_{z}\rangle(|\psi_{r}(0)\rangle)}
.\tag{S80}\label{S80}
\end{equation}
In this way, for an unknown initial state of the system, we can measure the change in atomic energy to obtain the instantaneous quantum Fisher information for the parameter $g$.
\nocite{*}

\bibliography{SupplementalMaterial}